\documentclass[draftclsnofoot, onecolumn, comsoc, 12pt]{IEEEtran} 
\usepackage[letterpaper, right=1in, left=1in, bottom=1in, top=1in]{geometry}
\usepackage{cite}
\usepackage[normalem]{ulem}
\usepackage{amsmath,amssymb,amsfonts}
\usepackage{graphicx}
\usepackage{textcomp}
\usepackage{xcolor}
\usepackage{amssymb}
\usepackage{amsmath}
\usepackage{cite}
\usepackage{stfloats}
\usepackage{epstopdf}
\usepackage{mdwlist}
\usepackage{algorithm}
\usepackage[noend]{algpseudocode}
\usepackage{tabularx} % for table of equations
\usepackage{orcidlink}
\usepackage{bbold}

\usepackage[all,2cell]{xy} \UseAllTwocells
\usepackage{booktabs}

\usepackage{tikz}
\usetikzlibrary{arrows.meta}
\usetikzlibrary{decorations.pathreplacing}
\usepackage{xcolor}
\usepackage{enumitem}
\usepackage{nicefrac}

\usepackage{subcaption}
\usepackage{dsfont}
\usepackage{mathrsfs}

\newtheorem{remark}{Remark}

\newtheorem{proof}{Proof}
\newtheorem{example}{Example}

\AtBeginDocument{}

\newcommand{\beq}{\begin{equation}}
	\newcommand{\eeq}{\end{equation}}
\newcommand{\beqa}{\begin{eqnarray}}
	\newcommand{\eeqa}{\end{eqnarray}}
\newcommand{\comment}[1]{}

\DeclareMathOperator*{\argmin}{arg\,min}

\usepackage{ulem}

\newif\ifbulletlist
\newif\iftext
\texttrue % comment to hide text

\usepackage{etoolbox}
\usepackage{lmodern}

\usepackage{tikz}% to draw at specific positions

\begin{document}

    \title{
		 Cooperative Learning-Based Framework \\for VNF Caching and Placement Optimization \\over Low Earth Orbit Satellite Networks
	}

\author{
	%(Work in progress)\\
	{\fontsize{13.35}{13}\selectfont Khai Doan,
	Marios Avgeris,
	Aris Leivadeas,
	Ioannis Lambadaris,
	and
	Wonjae Shin}
        \thanks{
		K. Doan and W. Shin are with Korea University, School of Electrical Engineering, Seoul, South Korea (e-mail:\{khaidoan, wjshin \}@korea.ac.kr).
	}
	\thanks{
		M. Avgeris and I. Lambadaris are with
		Carleton University, Department of Systems and Computer Engineering, Ottawa, ON, Canada
		(e-mail: \{mariosavgeris@cunet, ioannis@sce\}.carleton.ca). 
	}
	\thanks{
	  A. Leivadeas is with
		École de Technologie Supérieure, Department of Software and IT Engineering, Montreal, QC, Canada (e-mail: aris.leivadeas@etsmtl.ca).
	}
}

	%\markboth{
	%        ...}{
	%        ...=
	%...}
	
	\maketitle
	%\renewcommand{\baselinestretch}{1.36} \normalsize
	% The above \baselinestretch command does not change automatically.
	% A font size changing command should be executed to make the new value
	% in effect.  I trick this by using \normalsize, which does not change
	% the font size!
	%%%%%%%%%%%%%%%%%%%%%%%%%%%%%%%%%%%%%%%%%%%%%%%%%%%%%%%%%%%%%%%%%%%%%%
	
	\vspace{-1.5cm}
	
\begin{abstract}
Low Earth Orbit Satellite Networks (LSNs) are integral to supporting a broad range of modern applications, which are typically modeled as Service Function Chains (SFCs). Each SFC is composed of Virtual Network Functions (VNFs), where each VNF performs a specific task. In this work, we tackle two key challenges in deploying SFCs across an LSN. Firstly, we aim to optimize the long-term system performance by minimizing the average end-to-end SFC execution delay, given that each satellite comes with a pre-installed/cached subset of VNFs. To achieve optimal SFC placement, we formulate an offline Dynamic Programming (DP) equation. To overcome the challenges associated with DP, such as its complexity, the need for probability knowledge, and centralized decision-making, we put forth an online Multi-Agent Q-Learning (MAQL) solution. Our MAQL approach addresses convergence issues in the non-stationary LSN environment by enabling satellites to share learning parameters and update their Q-tables based on distinct rules for their selected actions. Secondly, to determine the optimal VNF subsets for satellite caching, we develop a Bayesian Optimization (BO)-based learning mechanism that operates both offline and continuously in the background during runtime. Extensive experiments demonstrate that our MAQL approach achieves near-optimal performance comparable to the DP model and significantly outperforms existing baselines. Moreover, the BO-based approach effectively enhances the request serving rate over time.
\end{abstract}
	
	
\section{Introduction}
In the evolving field of communication networks, satellite technologies have introduced innovative solutions for global connectivity. Low Earth Orbit Satellite Networks (LSNs) and space-air-ground integrated networks are gaining momentum for providing seamless global connectivity \cite{SHIN_2023, han2023adaptive}. The deployment of thousands of Low Earth Orbit (LEO) satellites by operators such as SpaceX and OneWeb represents a shift towards ultra-dense mega constellations aimed at delivering high-data-rate, high-capacity global communication services. As demand for real-time, complex applications grows—especially in remote and latency-sensitive areas—traditional standalone networks often fall short \cite{qin2023service}. Innovative network paradigms like Network Function Virtualization and Service Function Chains (SFCs) have been integrated into the context of LSNs as a solution \cite{he2023service}. Accordingly, Virtual Network Functions (VNFs) are designed and by chaining them in a specific order, SFCs are created to be delivered as the required network services. Leveraging the flexibility of virtualization allows for services to be deployed on satellite platforms to meet stringent requirements. This approach ensures seamless and robust service delivery, enhances reconfigurability, and reduces reliance on specialized hardware \cite{li2018service}.

While virtualization enables the distribution of SFC placements across the entire LSN, realizing its benefits is not straightforward. Effectively optimizing the placement of VNF components among the available satellites is crucial for enhancing the overall efficiency of the solution \cite{leivadeas2019vnf}. Specifically in the LSN context, traditional methods for SFC placement \cite{santos2022service} may face additional challenges in adapting to the unique characteristics of satellite networks, such as high mobility, complex time-varying topology, and resource competition. Reinforcement learning, and specifically Multi-Agent Reinforcement Learning (MAQL), has shown great potential in addressing these challenges \cite{KMAIW}. Reinforcement learning algorithms empower autonomous agents to develop optimal strategies through interactions with their environments and feedback in the form of rewards or penalties. For SFC placement in LSNs, where network conditions frequently change, MAQL enables multiple agents to collaboratively learn and adapt to these dynamic scenarios. Specifically, agents can dynamically optimize SFC placement by considering factors like satellite movement, evolving network conditions, and varying service demands. Additionally, MAQL's collaborative and decentralized approach enhances decision-making, increasing the adaptability and efficiency of SFC placement strategies in complex LSN environments. This method offers the potential for robust and adaptive SFC placement solutions tailored to the unique characteristics of LSNs.

%\comment{In this work, we put forth a MAQL method for SFC placement in LSNs, and then formulate the problem as a discrete-time stochastic control process aiming to maximize the system performance in the long run. As an extension of our previous work in \cite{KMAIW}, we delve deeper into the details of the proposed cooperative VNF/SFC placement method that tackles the dynamicity of the satellite network topology, followed by a formal algorithmic description. In addition, we convey a VNF caching policy that defines the optimal subsets of VNFs to be pre-installed on the satellites; the latter component complements the VNF placement functionality and further optimizes the system performance, in terms of maximizing the request serving ratio, given that satellite hardware capacity is limited. The main contribution of this work is fourfold:}

%The distinction between this work and the existing literature can be summarized as follows. 
%We consider a generic scenario where each satellite has a subset of VNFs pre-installed or cached. Given this constraint, our goal is to optimize the average end-to-end delay of the SFC deployment, over the long term. We formulate this problem within an optimal stochastic control framework, accounting for the dynamic nature of LSN topologies. 
In our previous work \cite{KMAIW}, we had formulated a Dynamic Programming (DP)-based solution and proposed a MAQL model for SFC deployment that optimized the long-term average service end-to-end delay. In this extension, we provide a detailed description of the proposed cooperative VNF/SFC placement method, followed by formal algorithmic definitions. In addition, considering the practical aspect of our approach—where each satellite has a subset of VNFs pre-installed/cached—we introduce a caching policy to determine the optimal VNF subsets for pre-installation on the satellites. This policy complements the placement mechanism by further optimizing performance and maximizing the request serving rate, given the constraints of satellite resources. %Since only cached or pre-installed VNFs can be placed and executed on a satellite, 
 Our contribution is fourfold:
\begin{enumerate}
    \item We refine the DP-based method from \cite{KMAIW} that aims to determine the optimal SFC placement policy to minimize end-to-end service delay. While this method provides an optimal solution, it is computationally intensive. Additionally, it necessitates centralized control and requires statistical information that is not always readily available.
    \item We revisit and refine the online MAQL approach from \cite{KMAIW}, which treats satellites as independent agents. This approach faces convergence challenges due to the non-stationary learning environment. To overcome this issue, we introduce a parameter sharing mechanism with distinct rules depending on satellites' actions, which enhances convergence and overall performance.
    \item To optimize the subsets of VNFs pre-installed on satellites, we propose a novel VNF caching policy based on Bayesian Optimization (BO). This approach iteratively refines installed VNF sets at satellites and continuously improves them in the background as the system operates. By integrating this caching strategy with our VNF placement approach, we create a comprehensive framework for LSN service deployment that enhances both end-to-end service delays and request serving rates.
    \item We experimentally demonstrate that our proposed framework closely approximates the optimal solutions. By leveraging the predictable movements of the satellites, our approach achieves improved request serving rates, end-to-end delays, and overall system cost. Additionally, we compare our method against those in the literature to highlight its effectiveness.
\end{enumerate}

The rest of this work is organized as follows: Section \ref{Sec:Related} discusses the related works.
%in the area of LSNs and VNF placement.
Section \ref{Sec:SystemModel} describes the system model. Section \ref{sec:problem-maql} provides the DP formulation and the proposed MAQL-based VNF placement solution. In Section \ref{sec:VNF_caching}, we introduce the BO-based VNF caching scheme. The experimental results are provided in Section \ref{sec:experiments} and finally a conclusion is drawn in Section \ref{sec:conclusion}.

\section{Related Works}
\label{Sec:Related}
Recent research has focused on exploiting the benefits of LSNs by optimizing the throughput and satellites' coverage to enhance the communication quality through beam scheduling, resource allocation and satellite formation design. For instance, in \cite{9832117} the authors proposed an Internet of Things (IoT) system that uses LEO satellites, emphasizing on its benefits such as low latency and broad coverage. Mokhtar \textit{et al.} \cite{898718}  provided throughput bounds and a beam scheduling algorithm for a satellite downlink system. The authors in \cite{9371230} used a unique 3D constellation optimization algorithm to optimize global coverage of an ultra-dense LEO satellite network. On the other hand, the authors in \cite{9351961} examined channel capacity in dense LEO multi-terminal satellite systems and explored the impact of satellite distribution and formation size on the capacity. Wang \textit{et al.} in \cite{9685355} addressed a multi-layer LEO satellite-terrestrial network, aiming to minimize satellite utilization. However, besides the communication aspects, the benefits of non-terrestrial networks, especially LSNs, can extend to computational services such as the proposed one.

\subsection{VNF Placement in Static Satellite Topologies}
LSNs create a breeding ground for diverse applications, including surveillance, tracking, mapping, weather forecasting, and disaster response \cite{KBKS2012}, which can be decomposed into various supporting tasks such as remote sensing and Earth observation. An emerging paradigm within this context involves breaking down application tasks and abstracting them into VNFs \cite{leivadeas2019vnf}. This approach aligns with the SFC paradigm, facilitating the deployment and interconnection of VNFs in LSNs, the intricate challenges of which have concerned the research community during the last few years. Gao \textit{et al.} \cite{GLTZM_2022} presented a hierarchical architecture for IoT users in satellite edge clouds, including IoT users, LEO satellites, and a cloud data center. They proposed a distributed VNF placement algorithm to minimize network bandwidth cost and service delay, considering satellite network characteristics like high latency and limited bandwidth. Also providing service to remote IoT users, the authors of \cite{GLK2022} aimed at optimizing satellite deployment costs by modeling the problem as a network payoff maximization problem, solved with decentralized allocation through a non-cooperative potential game.

\subsection{VNF Placement in Dynamic Satellite Topologies}
Qin \textit{et al.} \cite{qin2022sfc} examined a network model where service data is processed by satellites and sent to an Earth station. They formulated the problem as a congestion game to optimize delivery time and maintain continuous service, focusing on resource sharing and competition among SFCs. The authors then proposed two algorithms that achieve Nash Equilibrium to solve this problem. Jia \textit{et al.} \cite{JSLZH_2021} tackled a VNF-based service delivery problem involving a network operation control center, Geostationary Equatorial Orbit (GEO) satellites as SDN controllers, and LEO satellites. They proposed an approximation algorithm focused on efficient resource allocation for practical uses. By employing both types of GEO and LEO satellites, the study in \cite{JSMLGWD_2018} divided the network into a control plane and a data plane. A joint optimization approach for elastic network resource provisioning was then introduced, enabling adaptable VNF deployment and traffic routing. In \cite{CWZWG_2018}, the authors identified optimal SFC deployment paths that involve satellite deployment probabilities based on the satellite’s position and latitude. The authors of \cite{10241962} proposed a heuristic and a mixed-integer linear programming algorithm to minimize end-to-end service latency while maximizing the service acceptance rate.

Evidently, placing SFCs on satellite networks is a challenging problem; thus following the recent interdisciplinary trends,  Han \textit{et al.} \cite{han2023adaptive} devised an adaptive online SFC deployment algorithm for large-scale LSNs, based on deep reinforcement learning. Here, the authors proposed subnet division and a proximal policy optimization approach to reduce resource consumption in their solution. The cooperation/competition potential between individual satellites during the SFC placement problem was studied by Qin \textit{et al.} in \cite{qin2023service}; an approach based on potential game theory was adopted and three algorithms were proposed to minimize service delivery latency. On a different note, \cite{he2023service} suggested the use of a Tabu-search heuristic to solve the emerging integer non-linear programming problem of SFC deployment in satellite networks while minimizing VNF migrations. 

\subsection{Novelty of the Proposed Framework}
Despite the recent research interest around SFC deployment on LSNs, %the scopes of the these works do not account for the long-term system performance, and 
the majority of the works do not address the dynamic nature of the network nor optimize the long-term system performance. Additionally, the aforementioned works assume \textit{a priori} that satellites are capable of executing every type of VNF available in the system. However, in the LSN context, embedded satellite processor capability and storage are typically limited, hence, this is not a practical assumption. In the proposed framework, which extends our prior work in \cite{KMAIW}, we aim to specifically address these two limitations in the literature, by providing a holistic LSN service deployment framework (VNF placement and caching) that simultaneously optimizes the end-to-end service delay and the request serving rate.
    
\section{System Model} 
\label{Sec:SystemModel}
\subsection{Communication Model}
We assume an LSN that consists of multiple satellites, with $\mathbb{V} = \{v_1, \ldots, v_V \}$ being the set of their indices ($V = |\mathbb{V}|$). The satellites gather data (e.g., images of the Earth's surface) and cooperatively execute computational tasks for the supported applications. The satellites' movements are considered periodic, and also, we assume that an inter-satellite link (ISL) between a pair of satellites is active when the two satellites enter the communication range of each other. Additionally, each satellite $v \in \mathbb{V}$ is equipped with $R_v$ computational and $Z_v$ storage resources.
\begin{figure}
	\centering
	%\captionsetup{justification=centering}
	\includegraphics[scale=0.3]{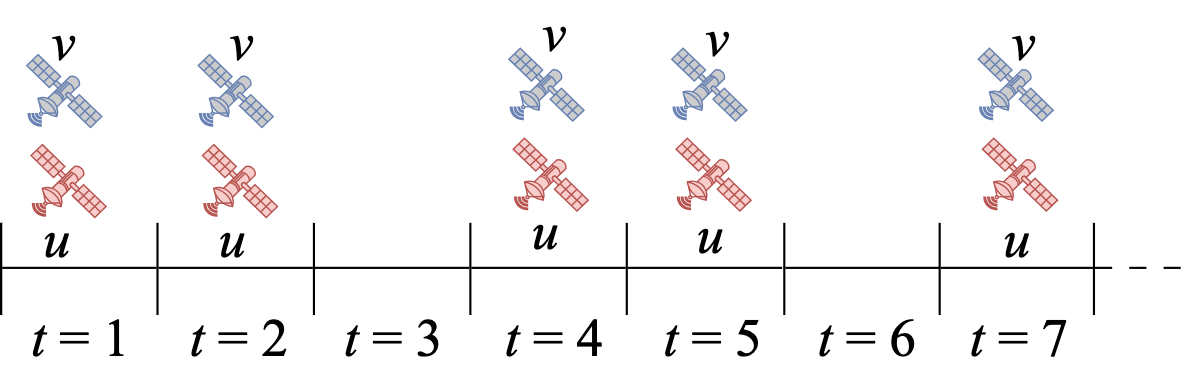}%example_satmobi.png
	\caption{Evolution of $(v,u)$-ISL in time, when $e_{v,u}\left(1\right)=1, \tau_{v,u}=2$, $T_{v,u}=3$. %An active ISL is portrayed with the presence of two satellites.
    Two satellites signify an active ISL.}
\label{fig:mobi_ex}
\end{figure}
    
The introduced satellite network operates in a slotted, infinite time horizon, where $t \in \mathbb{N}=\left\{1,2,\ldots\right\}$ denotes a time slot. We use the binary parameter $e_{v,u}(t) = 1$ (or $0$) to represent if the ISL between satellites $v$ and $u$, referred to as $(v, u)$-ISL, is active (or not). Due to the recurring nature of their orbits, satellites interact with each other at fixed intervals, and thus ISLs become active periodically as well. To model this periodicity, we introduce the following parameters; we assume that whenever the $(v,u)$-ISL is active, it remains active for $\tau_{v,u}$ time slots before becoming inactive. In addition, $T_{v,u}$ denotes the period of $(v,u)$-ISL (in time slots), implying that $T_{v,u} \ge \tau_{v,u}$ and $e_{v,u}\left(t\right) = e_{v,u}\left(t + T_{v,u}\right)$, $\forall t \in \mathbb{N}, \; v, u \in \mathbb{V}$. Furthermore, we denote by $T$ the period of the entire network connectivity which can be computed as the least common multiple of $T_{u,v},\forall u, v \in \mathbb{V}$. For simplicity, we assume that every data transfer between a satellite pair is completed within a single time slot using an inter-satellite free-space optical link with high transmission speed, long communication range, and high reliability. %We note that among the parameters describing a satellite network topology, only $T_{u,v}$ and $\tau_{u,v}$ are used in the proposed framework. 
In this model, we only focus on the the sets of parameters $T_{u,v}$ and $\tau_{u,v}$, $\forall v, u$ to describe the network topology and motions of satellites for simplifying the presentation and enhancing the clarify. An illustrative example is provided in Fig.~\ref{fig:mobi_ex} with the presence/absence of the two satellites in a time slot implying an active/inactive ISL. %Conversely, the absence of satellites signifies an inactive ISL. Additionally, $\tau_{v,u}=2$ suggests that once the $(v, u)$-ISL is active, it lasts for $2$ time slots; $T_{v,u} = 3$ represents the connection period meaning that the statuses of the $(v,u)$-ISL in time slot $t$ and $t+3$ are the same.

\subsection{SFCs, VNFs and Service Request Arrivals} 
\label{subsec:SFC}
As mentioned earlier, the satellites cooperatively execute computational tasks for certain types of applications. Each task requires data to be processed through the sequence of interconnected VNFs of an SFC. For instance, a wildfire detection involves several steps such as image processing, feature extraction, and fire classification \cite{KBKS2012}, each one carried out by a separate function represented by a VNF in this work. An ordered sequence of $l_h$ VNFs comprises an SFC $h\in \mathbb{H}$, where $\mathbb{H} = \left\{h_1, \ldots, h_H \right\}, $ is the index set of all the available SFCs/services, $H = |\mathbb{H}|$. Also, we let $\mathbb{F}=\{\mathcal{F}_1, \mathcal{F}_2,\ldots\}$ be the set of all available VNFs. We denote by $F^h_f \in \mathbb{F}$ the $f^{\text{th}}$ VNF of SFC $h$ and by $\mathbb{S}_h = \left\{ F^h_f \; | \; f \in \{1,...,l_h\} \right\}$ the set of VNFs comprising SFC $h\in\mathbb{H}$. For a satellite to execute a specific VNF, the corresponding source code needs to be installed in the satellite's computing system. The storage used for caching is excluded from the $Z_v$ storage resources of a satellite.  We denote by $\mathbb{F}_v$ the pre-installed set of VNF types on satellite $v\in \mathbb{V}$, with $|\mathbb{F}_v|$ being its caching capacity. A VNF $F^h_f$ can only be placed/executed on satellite $v$ if $F^h_f\in \mathbb{F}_v$. Due to their limited physical resources, it is not always feasible to install every available VNFs on the satellites. Therefore, we consider a scenario where a subset of VNF images has been pre-installed on each satellite. This means that $\mathbb{F}_v \subseteq \mathbb{F}$, covering the unique case $\mathbb{F}_v = \mathbb{F}$ where $v \in \mathbb{V}$.

At the end of time slot $t$, a \textit{service request} is initiated with a probability $\mu$. A request initiation triggers satellite $v \in \mathbb{V}$ to request the placement of an SFC $h \in \mathbb{H}$. We call this satellite the \textit{requester}; $\mu^r_v$ and $\mu^s_h$ are the probability that satellite $v$ will be the requester and the probability that SFC $h$ will be requested, respectively, conditioned on the occurrence of a service request. A service request for SFC $h$ will expire at the end of time slot $t + D_h$ if not successfully served, where $D_h$ is the SFC's end-to-end delay tolerance. \comment{To present the DP and learning-based solutions \cite{Sutton1998}, we assume up to one incoming SFC request per time slot \footnote{For cases with multiple simultaneous requests, they can be buffered to eliminate spatial relations; then, each one is processed in a time slot. Thus, the proposed frameworks remain applicable.}; however, the execution of an SFC can span multiple time slots. Consequently, a satellite may handle multiple requests within a given time slot. These requests can include those forwarded from other satellites and those initiated by the satellite itself.
In this system model, we assume up to one SFC request initiated per time slot. For cases where a satellite initiates multiple requests simultaneously, these requests can be buffered to eliminate their temporal relation. Then, each one is processed in a time slot and the proposed framework in this paper remains applicable. In addition, since the execution of an SFC can span multiple time slots, a satellite can handle multiple requests within a given time slot including requests of its own and those forwards from other satellites. In this work, each request is associated with a timestamp indicating its initiation time, and a satellite will handle requests in the ascending order of their initiation time. Timestamps will be formally defined in Section \ref{sec:learning_model}. The concept of buffer and timestamps described above help determining the order of requests to be processed; hence, preventing resource conflicts at satellites.} 
In our system model, we assume that at most one SFC request is initiated per time slot. Additionally, since the execution of an SFC may span multiple time slots, a satellite can manage multiple requests within a time slot, including its own and those forwarded from other satellites. In this work, each request is assigned a timestamp (described in Section \ref{sec:learning_model}) corresponding to its initiation time, and requests are processed in ascending order of these timestamps. In a practical context where a satellite initiates or receives multiple requests with the same initiation time slot, these requests can be buffered to eliminate their temporal overlap. Each buffered request is processed in a time slot, and the proposed framework in this paper remains applicable. Timestamps along with the use of buffers help determine the order in which requests are processed, thereby preventing resource conflicts.

\subsection{VNF Execution and Inter-Satellite Data Transfer}
\label{subsec:execute_SFC}
We use the terms \textit{placement} and \textit{execution} interchangeably throughout this work. Executing SFC $h$ means that all the VNFs $F^h_f$ for $f=1,\ldots,l_h$ are executed in the given order. That means the output of the $f^{\text{th}}$ VNF becomes the input of the $(f+1)^{\text{th}}$ VNF in the sequence. The output of the last VNF in the sequence bears the results carried out by SFC $h$ and needs to be returned to the requester satellite to mark the request as successfully served. We also denote by $q^h_f$ and $g^h_f$ the computational and storage resources required to execute $F^h_f \in \mathbb{S}_h$ and store its output.

We assume that if satellite $v$ executes VNF $F^h_f$, and VNF $F^h_{f+1}$ is assigned to another satellite $u$ following a placement policy (e.g., because $F^h_{f+1}$ has not been pre-installed on $v$), $v$ needs to store the output of $F^h_f$ in its storage before forwarding\footnote{The forwarding path may consist of multiple satellites as relays.} it to satellite $u$. In this case, when $F^h_f$ has been executed, $q^h_f$ computational resources on satellite $v$ are released, but $g^h_f$ storage resources are consumed to temporarily store its output. That happens because the communication can occur only when the $(v,u)$-ISL is active. After forwarding the data, $g^h_f$ storage resources are released on satellite $v$, while $g^h_f$ storage resources are consumed on satellite $u$ to store the received data. That means that data forwarding might fail if the receiving satellite does not have sufficient available storage space. On the contrary, if a satellite is executing a sequence of VNFs, the intermediate data is not stored in its storage. 

\begin{figure}
	\centering
	%\captionsetup{justification=centering}
	\includegraphics[scale=0.3]{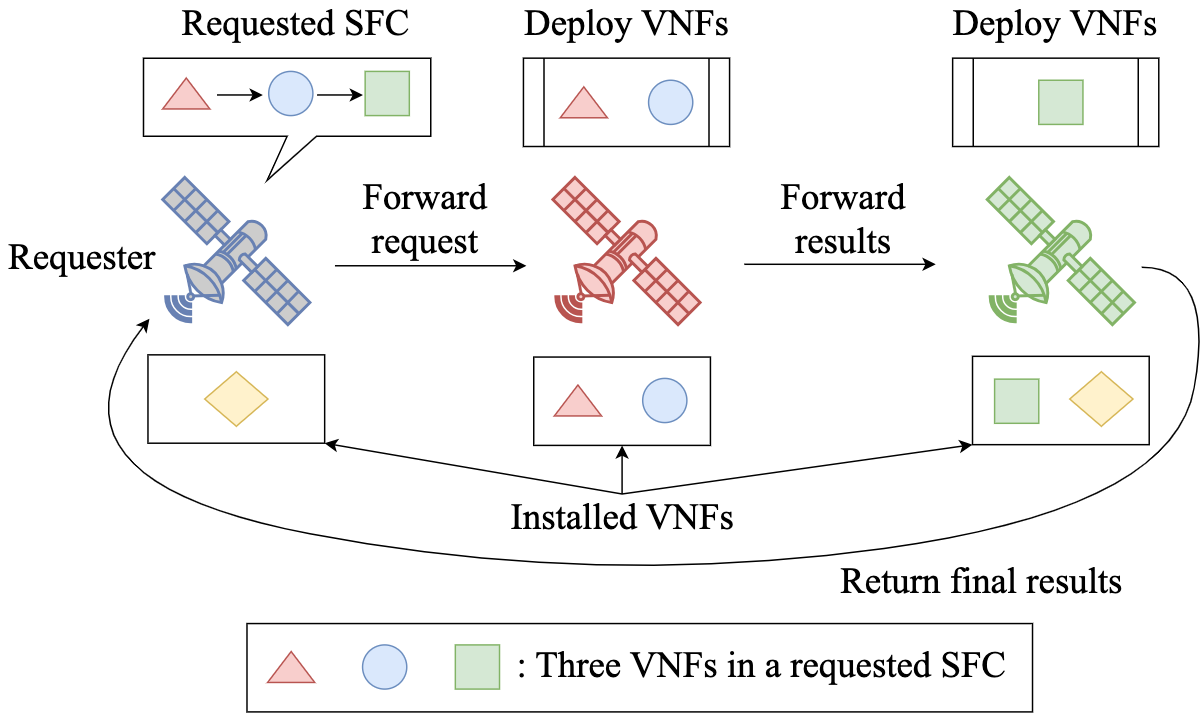}%example_satmobi.png
	\caption{The VNF/service placement problem on an LSN.} 
    \label{fig:concep_ex}
\end{figure}

%Given that a VNF $F^h_f\in \mathbb{F}_v$ that is executed on satellite $v$ finishes after $d^h_{f,v}$ time, and that SFC $h$ has an end-to-end delay tolerance equal to $D_h$, the problem of calculating a VNF placement on the available satellites that satisfies the delay constraints emerges. An illustration of the problem is given in Fig.~\ref{fig:concep_ex}. We address two sub-problems in this work:
In this work, we address the following two sub-problems:
\begin{itemize}
    \item The first sub-problem is to assign VNFs to satellites and define data routing paths to optimize the overall system performance. We refer to this as the \textit{VNF/service placement} problem which is formally defined and solved in Section \ref{sec:problem-maql}.
    %The first sub-problem involves determining which satellite will execute which VNF and when, establishing in this way the most efficient data routing path among them, in terms of optimizing system performance while respecting resource constraints and Service Level Agreements. We refer to this problem as the \textit{VNF/service placement} problem and it is formally defined and solved in Section \ref{sec:problem-maql}. 
    \item The second sub-problem is to define the optimal VNF subset installed on each satellite. We refer to this as the \textit{VNF caching} problem which is formally defined and solved in Section \ref{sec:VNF_caching}.
    %Given that the satellites are limited on what VNFs they can execute based on their pre-installed set, a second sub-problem arises from the need to discover the optimal VNF subset for each satellite. We refer to this problem as the \textit{VNF caching} problem and it is formally defined and solved in Section \ref{sec:VNF_caching}.
\end{itemize}
Fig.~\ref{fig:concep_ex} illustrates a VNF placement constrained by installed VNF subsets. Table \ref{table:notation} summarizes the main notation. To reduce complexity, we will gradually augment the notation with additional variables as needed. In addition, we note that service requests and ISL transitions occur in each time slot. The former, along with the available resources of each satellite, constitutes the system state, which will be defined in the next section. The latter is due to satellite motions. Therefore, a time slot can be interpreted as a unit of time that quantifies both components.

\begin{table}[H]
    \centering
    \caption{Summary of main notation.}
    \label{table:notation}
    \begin{tabular}{|c|c|}%{\columnwidth}{|c|X|}
    \hline
    \textbf{Notation} & \textbf{Description} \\
    \hline
    \(\mathbb{V} = \{v_1, \ldots, v_V \}\) & Set of LSN's satellites \\
    \hline
    \(e_{v,u}(t)\) & $(v,u)$-ISL state at time slot $t\in \mathbb{N}$, $ u, v \in \mathbb{V}$ \\
    \hline
    \( \tau_{v,u}, T_{v,u}, T \) & Active  ISL duration, period,  system period \\
    \hline
    \( R_v, Z_v \) & Computational, storage capacity of $v \in \mathbb{V}$ \\
    \hline
    \(\mathbb{H} = \left\{1, \ldots, H \right\} \) & Set of LSN's available SFCs/services  \\
    \hline
    \(\mathbb{S}_h = \{F^h_f \} \) & Set of VNFs comprising SFC $h\in\mathbb{H}$, with $f \leq l_h$ ($l_h$ being SFC's $h$ length) \\
    \hline
    \(\mathbb{F}_v = \{\mathcal{F}_i \} \) & Set of cached VNF types on satellite $v\in\mathbb{V}$, $i\leq |\mathbb{F}_v|$ \\
    \hline
    \( \mu, \mu^r_v, \mu^s_h \) & Service request, requester satellite, $v\in\mathbb{V}$, and requested SFC, $h \in \mathbb{H}$, probabilities \\
    \hline
    \( D_{h} \) & SFC $h \in \mathbb{H}$ end-to-end delay tolerance \\
    \hline
    \( q^h_f, g^h_f \) & $F^h_f$ computational, storage requirements \\
    \hline
    \( d^h_{f,v} \) & $F^h_f$ execution delay on $v\in\mathbb{V}$ \\
    \hline
    \( \mathbf{p} = \left\{\boldsymbol{\phi}, \mathbf{m} \right\} \) & (DP) SFC placement \\
    \hline
    \( \boldsymbol{\phi} = \{\phi_{k} \} \) & Set of satellites $\phi_{k} \in \mathbb{V} $ that handle the request at each time slot $k \in \mathbb{N}$ \\
    \hline
    \( \mathbf{m} = \{m_f\}  \) & Set of time slots  $m_f \in \mathbb{N}$ where VNF $F^h_f \in\mathbb{F}_v$ is activated \\
    \hline
    \( K \) & Largest end-to-end SFC delay tolerance\\
    \hline
    \( \mathbf{r}_v = \{r_{v,k}\} \) & Computational resource of $v \!\in \!\mathbb{V}, \forall k \!\leq \!K$\! \\
    \hline
    \( \mathbf{z}_v = \{z_{v,k}\} \) & Storage resource of $v \in \mathbb{V}, \forall k \leq K$ \\
    \hline
    \( \mathbf{x} = \{\mathbf{x}_v, h, n \} \) & (DP) System state: $\mathbf{x}_v = (\mathbf{r}_v, \mathbf{z}_v)$, $\forall v \in \mathbb{V}$, requested SFC $h \in \mathbb{H}$, requester  $n \in \mathbb{V}$
    \\
    \hline
    \( \tilde{\mathbf{x}} = \{\tilde{\mathbf{x}}_v, \tilde{h}, \tilde{n} \} \) & (DP) Transitioned system state  \\
    \hline
    \(\mathcal{P}\{\mathbf{x} \rightarrow_\mathbf{p}\tilde{\mathbf{x}} \}\) & (DP) State transition probability given $\mathbf{p}$\\
    \hline
    \( \mathcal{C}\left(\mathbf{x}, \mathbf{p} \right)\) & (DP) Cost of placement $\mathbf{p}$ on state $\mathbf{x}$\\
    \hline
    $\mathbb{Y}_v$ & Set of requests in the buffer of satellite $v$\\
    \hline
    \( \mathbf{y} = \left\{h, n, f, \hat{t} \right\} \)  & (MAQL) Request handled by satellite $v\in\mathbb{V}$, $h \in \mathbb{H}$, $n \in \mathbb{V}$, $\hat{t} \in \mathbb{N}$, $\mathbf{y}\in\mathbb{Y}_v$  \\
    \hline
     \( \mathbf{s}_v = \{(r_v, z_v), \mathbf{y}\} \) & (MAQL) Satellite $v\in\mathbb{V}$ state, $\forall\mathbf{y}\in\mathbb{Y}_v$\\
    \hline
    \( \mathbf{a}_v = \{a\} \) & (MAQL) Satellite $v\in\mathbb{V}$ actions, $\forall \mathbf{y}\in\mathbb{Y}_v $ \\
    \hline
    \( c_v(\mathbf{y}, a) \) & (MAQL) Cost when satellite $v\in\mathbb{V}$ performs action $a\in \mathbf{a}_v$ on handled request $\mathbf{y}\in\mathbb{Y}_v$\\
    \hline
    $\mathbb{A}_v(\mathbf{y})$ & The set of available actions for request $\mathbf{y}$ of satellite $v$ \\
    \hline
    \( \pi_v\left(\mathbf{y}\right) \rightarrow a \)  & (MAQL) Placement policy $v\in\mathbb{V}, a \in \mathbb{A}_v(\mathbf{y})$\\
    \hline
    \( C_v\left(\mathbf{s}_v, \mathbf{a}_v\right) \) & (MAQL) Total cost of satellite $v\in\mathbb{V}$ for state $\mathbf{s}_v$ and action sequence $\mathbf{a}_v$ \\
    \hline
    $\zeta$ & Order of  current time slot in system period\\
    \hline
    $\mathbb{Q}_v$ & Q-table of satellite $v$\\
    \hline
    \( Q_v^{\zeta}\left(\mathbf{y}, a\right) \) & Q-value $\in \mathbb{Q}_v,\mathbf{y}\in\mathbb{Y}_v, a\in \mathbf{a}_v, v\in\mathbb{V}$, and $\zeta \in \{1,\ldots,T \}$ \\
    \hline
    \(\lambda, \delta\) & Learning rate and discount factor \\
    \hline
    \( \boldsymbol\theta \) & System-wide VNF caching strategy \\
    \hline
    \( M_{\Pi}(\boldsymbol\theta) \) & Request serving rate under policy $\Pi$ and caching strategy $\boldsymbol\theta\in \boldsymbol{\Theta}$ \\
    \hline
    \( \dot\mu(\boldsymbol\theta), \dot\sigma(\boldsymbol\theta), \mathcal{K}(\boldsymbol\theta, \tilde{\boldsymbol\theta}) \) & Predicted mean, standard deviation, kernel function for caching strategies $\boldsymbol\theta, \tilde{\boldsymbol\theta}\in \boldsymbol{\Theta}$ \\
    \hline
    \( I(\boldsymbol\theta) \) & Acquisition function for caching strategy $\boldsymbol\theta$ \\
    \hline
\end{tabular}
\end{table}

\section{VNF Placement in LSNs}
\label{sec:problem-maql}
\subsection{Dynamic Programming Formulation}
\label{sec:DP} 
The execution of an SFC can span several time slots, and its VNFs can be activated in multiple satellites at different time slots sequentially. We define the service placement as a tuple that contains both the indices specifying which satellites will handle data relaying and which will execute VNFs of the requested SFC, as well as their activation time slots. We denote it as $\mathbf{p} = \left\{\boldsymbol{\phi}, \mathbf{m} \right\}$, where:
\begin{equation} 
    \label{eq:placement1}
    \boldsymbol{\phi} = \{\phi_{k} \;|\; \phi_{k} \in \mathbb{V}, k \in \mathbb{N} \},
\end{equation}
\noindent and $\phi_{k}$ denotes the index of the satellite that handles the request at the $k^{\text{th}}$ time slot, with $k=1$ representing the time slot when the placement $\mathbf{p}$ is performed. $\phi_1$ and $\phi_{|\boldsymbol{\phi}|}$ are, respectively, the indices of the requester and the last satellite handling the request before the final result is obtained by the requester. We note that, by ``handling'' we refer to one of the followings: forwarding, carrying, rejecting a request or executing a VNF in it. Thus, calculating the number of elements in $\boldsymbol{\phi}$, i.e., $|\boldsymbol{\phi}|$, gives us the end-to-end SFC/service execution and data transfer time, in time slots, for placement $\mathbf{p}$. Set $\mathbf{m}$ is defined as:
\begin{align}
    \label{eq:placement2}
    \mathbf{m} = \{m_f \;|\; m_f \in \mathbb{N},  f \in \{1, \ldots, l_h\}\},
\end{align}
where $m_f$ is the time slot that VNF $F^h_f \in\mathbb{F}_v$ is activated, with $m_f=1$ representing again the time slot when the placement $\mathbf{p}$ is performed. A request rejection is represented by $\mathbf{p} = \emptyset$. A simple and intuitive example follows:
\begin{example} \label{ex:placement}
    The requester satellite $1$ requests SFC $1$ comprised of two VNFs, $F_1^1$ and $F_2^1$. We assume for simplicity that $\mathbb{F}_1 = \{F_1^1 \}$ and $\mathbb{F}_2 = \{ F_2^1\}$, i.e., the first and second VNFs are cached on satellites $1$ and $2$, respectively. It is also assumed that each satellite requires one time slot to execute its VNF. A feasible service placement is:
    \begin{itemize} 
        \item $t=1$: Satellite $1$ executes $F^1_1$.
        \item $t=2$: Satellite $1$ forwards its output to satellite $2$.
        \item $t=3$: Satellite $2$ receives the data and executes $F^1_2$.
        \item $t=4$: Satellite $2$ forwards the results back to satellite $1$, the requester.
    \end{itemize}
    This placement $\mathbf{p}$ spans $4$ time slots. From definition, we have that $\boldsymbol{\phi} = \left\{1, 1, 2, 2 \right\}$, and $\mathbf{m} = \left\{1, 3\right\}$, as the two VNFs are deployed at $t=1$ and $t=3$, respectively. $\square$
\end{example}

We note that when a placement $\mathbf{p}$ is made at time slot $t$, the required resources of all involved satellites in future time slots are reserved accordingly. We define $K = \underset{h \in \mathbb{H}}{\max}~D_h$ as the largest end-to-end delay tolerance among all supported SFCs. We represent the system's state by $\mathbf{x}$ which is updated at the beginning of every time slot; $\mathbf{x}$ encapsulates the satellites' available resources for up to $K$ future time slots, the index of the requested SFC, as well as the index of the requester satellite.\footnote{It is sufficient for $\mathbf{x}$ to convey $K$ time slots from the current one because no placement can span more than $K$ time slots.} The vector of available resources of satellite $v$ for the next $K$ time slots is given by  $\mathbf{x}_v = (\mathbf{r}_v, \mathbf{z}_v)$, with $\mathbf{r}_v = \{r_{v,k} \;|\; r_{v,k} \leq R_v, \forall k \in \{1,...,K\}\}, ~\mathbf{z}_v = \{z_{v,k} \;|\; z_{v,k} \leq Z_v, \forall k \in \{1,...,K\}\}$, where $r_{v,k}$ and $z_{v,k}$ are the available computational  and storage resources of satellite $v$, respectively, at time slot $k$. These allow us to formally define the system state as:
\begin{align}
    \label{eq:system_state}
    \mathbf{x} = \{\mathbf{x}_v, h, n \;|\; \forall v \in \mathbb{V}, h \in \mathbb{H}, n \in \mathbb{V} \},
\end{align}
where $h$ and $n$ are the indices of the requested SFC and the requester at the end of the previous time slot, respectively. If there is no request in the previous time slot, $h = n = 0$.

We define $\tilde{\mathbf{x}} = \{\tilde{\mathbf{x}}_v, \tilde{h}, \tilde{n} \;|\; \forall v \in \mathbb{V},\tilde{h} \in \mathbb{H}, n \in \mathbb{V} \}$ as the transitioned system state resulted by placement $\mathbf{p}$ in the beginning of the following time slot; $\tilde{\mathbf{x}}_v = (\tilde{\mathbf{r}}_v, \tilde{\mathbf{z}}_v)$ represents the transitioned vector of available resources on satellite $v$. We compute the components of $\mathbf{x}_v$ as follows:
\begin{enumerate}
    \item[(i)] if $v = \phi_{m_{f}} \in \boldsymbol{\phi}$ and  $m_{f} \in \mathbf{m}$, meaning that satellite $v$ is executing VNF $F^h_f$, hence, 
    $\phi_{m_f + m} = v, \; \forall m \in [0, d^h_{f,v}-1]$. Therefore, $\tilde{r}_{v,k} = r_{v,\left(k+1\right)} - q^h_{f}$, and  $\tilde{z}_{v,k} = z_{v,\left(k+1\right)}$, $\forall k \in [m_f, m_f+d^h_{f,v}-1]$.
    \item[(ii)] else if $v = \phi_m \in \boldsymbol{\phi}$, $m \notin \mathbf{m}$, and $\exists m_f \in \mathbf{m}$ such that either $m_f < m < m_{f+1}$ or $m > m_{f=l_h}$, meaning $v$ stores the output of $F^h_f$. Therefore, $\tilde{r}_{v,(m-1)} = r_{v,m}$ and $\tilde{z}_{v,(m-1)} = z_{v,m} - g^h_{f}$. We note that if $m_f < m$, with $m_f \in \mathbf{m}$, then $m \ge 2$.
    \item[(iii)] else either $v \notin \boldsymbol{\phi}$, i.e., satellite $v$ is not part of the placement $\mathbf{p}$, then, $\tilde{r}_{v,k} = r_{v,\left(k+1\right)}, \tilde{z}_{v,k} = z_{v,\left(k+1\right)}$, $\forall k \in [1, K-1]$, or the placement $\mathbf{p}$ spans less than $K$ time slots ($|\boldsymbol{\phi}| < K$), then, $\tilde{r}_{v,k} = r_{v,\left(k+1\right)}, \tilde{z}_{v,k} = z_{v,\left(k+1\right)}$, $\forall k \in [|\boldsymbol{\phi}|+1, K-1]$.
\end{enumerate}
Since a valid placement $\mathbf{p}$ cannot span more than $K$ time slots, $\tilde{r}_{v,K} = R_v$, and $\tilde{z}_{v,K} = Z_v$ always. A representative example is provided below to aid the understanding of the definitions of system states and state transition: 
\begin{example}
    Let us consider an LSN consisting of two satellites, $V=2$, where each one is equipped with computational resources equal to $R_1 = R_2 = 2$ and storage resources equal to $Z_1 = Z_2 = 3$. Two services are offered by the system as SFCs, $H=2$, of length $l_1 = l_2 = 2$ and maximum end-to-end delay tolerances equal to $D_1 = 6$ and $D_2 = 7$ time slots. We have $K=\max(D_1,D_2)=7$. The execution of each of their VNF components consumes $q^1_f = q^2_f = 1$ computational resources and requires $d^1_{f,v} = d^2_{f,v} = 1$, $v \in \mathbb{V}$, time slots to complete. Their output consumes $g^1_f = g^2_f = 1$ storage resources when stored. Data forwarding takes $1$ time slot. Let $T_{1,2}=T=2$ and $\tau_{1,2}=1$, i.e, the two satellites are in the communication range of each other every $2$ time slots and once they are connected, their connection remains active for $1$ time slot. Initially, we assume that the two satellites are not connected, i.e., the $(1,2)$-ISL is inactive, thus $e_{1,2}(1) = 0$, while they have all their resources available, i.e., $\mathbf{r}_v = \{r_{v,k}=2|k=1,\ldots,7\}$ and $\mathbf{z}_v =\{z_{v,k}=3|k=1,\ldots,7\}$ for $v=1,2$. Additionally, let $\mathbb{F}_1 = \{F_2^1 \}$ and $\mathbb{F}_2 = \{F_1^1 \}$, i.e., satellites 1 and 2 come pre-installed with the second and the first VNF of SFC $1$, respectively.

\captionsetup[figure]{
  %labelformat=simple,
  %labelsep=colon,
  textfont=normal, % Caption text will not be bold
  %justification=raggedright,
  labelfont={color=black}, % Label will be blue and bold
  skip=10pt % Adjust vertical space if needed
}
\begin{figure}
	\centering
	%\captionsetup{justification=centering}
	\includegraphics[scale=0.44]{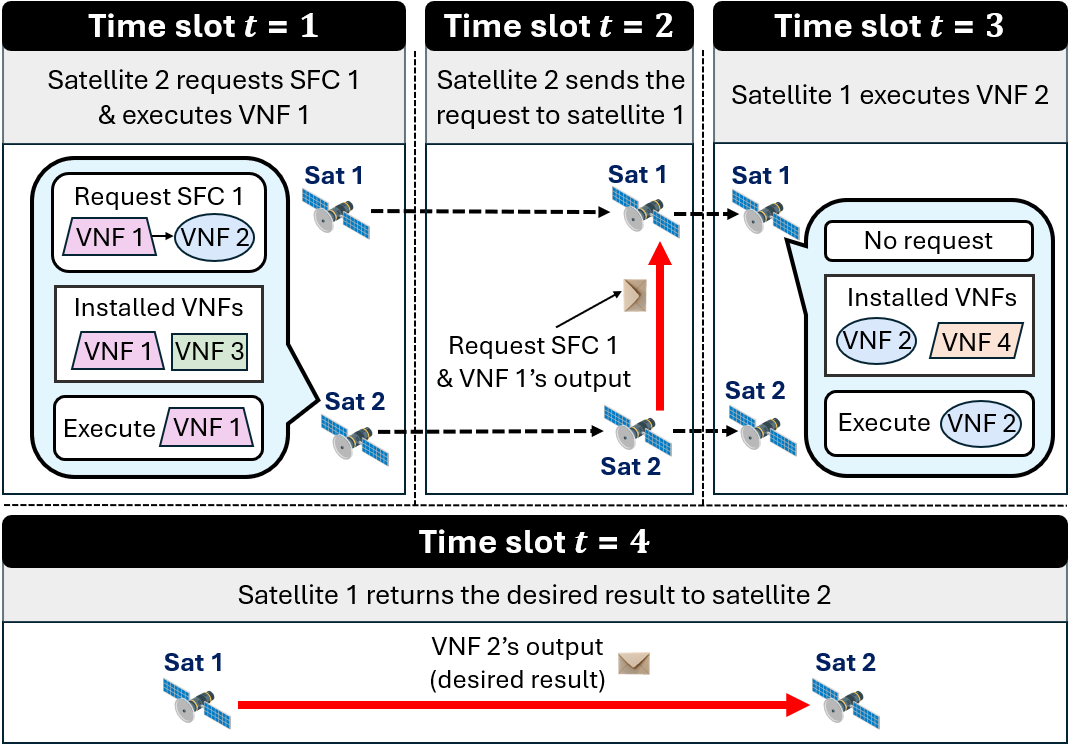}
	\caption{Visualization of SFC deployment in Example 2.} 
\label{fig:example2_illustrate}
\end{figure}

    At the end of time slot $t=1$, the requester satellite $n=2$ initiates a service request for SFC $h=1$. Then, the system state at the beginning of time slot $t=2$ is $\mathbf{x} = \{\mathbf{x}_1, \mathbf{x}_2, 1, 2 \}$, where $\mathbf{x}_1 = \{\mathbf{r}_1, \mathbf{z}_1 \}$ and $\mathbf{x}_2 = \{\mathbf{r}_2, \mathbf{z}_2 \}$. We visualize a feasible service placement of this example in Fig.~\ref{fig:example2_illustrate}, with the following details:
    \begin{itemize}
        \item \!$t=1$: Satellite 2 executes $F_1^1$ which has been installed.
        \item \!$t=2$: Satellite 2 forwards its output to satellite 1.
        \item \!$t=3$: Satellite 1 receives the data and executes $F_2^1$.
        \item \!$t=4$: Satellite 1 forwards the output of $F_2^1$ \!(the desired\!\! result) back to satellite 2, the requester.
        % \item $t=6$: satellite 2 receives the desired result and the process terminates.
    \end{itemize}
    From this process, we obtain $\boldsymbol{\phi} = \{2, 2, 1, 1\}$ and $\mathbf{m} = \{1, 3 \}$. We can calculate the components of the transitioned vectors of available resources $\tilde{\mathbf{x}}_1$, $\tilde{\mathbf{x}}_2$, through the following steps:
        \begin{itemize}
            \item For $v = \phi_{m_1} = 2$ with $m_1 = 1$, condition (i) is satisfied. The same stands for $v = \phi_{m_2} = 1$ with $m_2 = 3$. Therefore, $\tilde{r}_{1,2} = r_{1,3} - 1 = 2$ and $\tilde{z}_{1,2} = z_{1,3} = 3$.
            \item For $v=\phi_2=2, \; m = 2 \notin \mathbf{m}$ and $\exists m_1 \in \mathbf{m}$ such that $m_1=1 < m < m_2=3$, condition (ii) is satisfied. The same stands for $v=\phi_4=1$. Therefore, $\tilde{r}_{2,1} = r_{2,2} = 2$ and $\tilde{z}_{2,1} = z_{2,2} - 1 = 2$. Also $\tilde{r}_{1,3} = r_{1,4} = 2$ and $\tilde{z}_{1,3} = z_{1,4} - 1 = 2$.
            %Next, we define $\tilde{r}_{vk}$ and $\tilde{z}_{vk}$ for $v=1,2$ and $k=1,3$. 
            % We see that $(v=2, m=2)$ and $(v=1, m=4)$ satisfy the conditions in (ii), therefore, $\tilde{r}_{21} = r_{22} = 2$ and $\tilde{z}_{21} = z_{22} - 1 = 2$ ($g_1^1 =1$ in this example); $\tilde{r}_{13} = r_{14} = 2$ and $\tilde{z}_{13} = z_{14} - 1 = 2$ ($g_2^1=1$ in this example).
            \item Since the placement $\mathbf{p}$ spans $|\boldsymbol{\phi}|=4<K=7$ time slots, condition (iii) is satisfied for calculating the rest of the components. Therefore, $\tilde{r}_{v,k} = r_{v,k}$ and $\tilde{z}_{v,k} = z_{v,k}$, $\forall v\in\{1,2\}, \; \forall k \in \{5, 6\}$. Finally, $\tilde{r}_{1,7} = \tilde{r}_{2,7} = 2$ and $\tilde{z}_{1,7} = \tilde{z}_{2,7} = 3$. $\square$
            %It remains to define $\tilde{r}_{vk}$ and $\tilde{z}_{vk}$ for $v=1,2$ and $k=4, \ldots, 7$. 
            %Since the placement $\mathbf{p}$ spans 5 time slots ($|\mathbf{u}|=5$), $\tilde{r}_{vk} = r_{vk}$ and $\tilde{z}_{vk} = z_{vk}$ for $v=1,2$ and $k=4, 5, 6$. Finally, $\tilde{r}_{vK} = R_v$ and $\tilde{z}_{vK}=Z_v$.
        \end{itemize} 
\end{example}
Next, we define the transition probability from $\mathbf{x}$ to $\tilde{\mathbf{x}}$, given a placement $\mathbf{p}$, as:
\begin{align}\label{eq:trans_prob}
    \mathcal{P}\left\{\mathbf{x} \underset{\mathbf{p}}{\rightarrow} \tilde{\mathbf{x}} \right\} =
    \begin{cases}
        \mu \times \mu^r_{\tilde{n}} \times \mu^s_{\tilde{h}} &,  \text{if a request arrives}, \\
        1- \mu &,  \text{otherwise}.
    \end{cases}
\end{align}
Additionally, we define the system cost $\mathcal{C}\left(\mathbf{x}, \mathbf{p} \right)$ for performing placement $\mathbf{p}$, while on state $\mathbf{x}$, as: $\mathcal{C}(\mathbf{x}, \mathbf{p}) = |\boldsymbol{\phi}|$ if the request is served, and $\mathcal{C}(\mathbf{x}, \mathbf{p}) = C_p \gg 0$ if rejected.

To accommodate the calculation of the long-term, time-dependent discounted cost, we augment the state and placement variables with time slot notation, $\mathbf{x}(t)$ and $\mathbf{p}(t)$, respectively. Then, this discounted cost can be given by:
\begin{align}\label{eq:longterm_cost_DP}
    \sum_{t=1}^{\infty} \gamma^t \mathbb{E}\left[\mathcal{C}\left(\mathbf{x}\left(t\right), \mathbf{p}\left(t\right) \right)\right],
\end{align}
where $\gamma \in \left[0, 1\right]$ is the discount factor. The objective of the VNF placement task is to minimize the discounted cost (\ref{eq:longterm_cost_DP}). Let $J\left(\mathbf{x}\right)$ be the minimum discounted cost which can be calculated through the following DP Equation:
\begin{align}\label{eq:DP}
    J\left(\mathbf{x}\right) = \underset{\mathbf{p}}{\min}\left\{\mathcal{C}\left(\mathbf{x}, \mathbf{p} \right) + \gamma \; \mathcal{P}\left\{\mathbf{x} \underset{\mathbf{p}}{\rightarrow} \tilde{\mathbf{x}} \right\} J\left(\tilde{\mathbf{x}}\right) \right\}.
\end{align}
Solving \eqref{eq:DP} produces the optimal service placement $\mathbf{p}$ that minimizes the end-to-end delay for the requested service. 

A valid placement $\mathbf{p}$ adheres to the system constraints. We remind that $\mathbf{p} = \left\{\boldsymbol{\phi}, \mathbf{m} \right\}$, with $\boldsymbol{\phi}$ and $\mathbf{m}$ having been defined in Eqs. (\ref{eq:placement1}) and (\ref{eq:placement2}) respectively. $\phi_k \in \boldsymbol{\phi}$ denotes the index of the satellite handling the request in the $k^{\text{th}}$ time slot and $m_f \in \mathbf{m}$ denotes the time slot at which the $f^{\text{th}}$ VNF is executed; $m_f=1$ denotes the current time slot. Therefore, $\phi_{m_f}$ is the index of the satellite executing the $f^{\text{th}}$ VNF. Then, the set of VNF placement constraints with respect to (wrt.) the requested SFC $h$ is given as:
%is subjected to a set of constraints that will be introduced hereafter. Due to the notation complexity, we would like to recall that $\mathbf{p} = \left\{\boldsymbol{\phi}, \mathbf{m} \right\}$ with $\boldsymbol{\phi}$ and $\mathbf{m}$ are defined in Eqs. (\ref{eq:placement1}) and (\ref{eq:placement2}), respectively. $\phi_k$ denotes the index of the satellite handling the request in the $k^{\text{th}}$ time slot and $m_f$ denotes the time slot at which the $f^{\text{th}}$ VNF is executed, where the current time slot is considered as the first one. Therefore, $\phi_{m_f}$ is the index of the satellite executing the $f^{\text{th}}$ VNF. Then, the set of VNF placement constraints with respect to the requested SFC $h$ is as follows:}
\begin{align}
    &|\boldsymbol{\phi}| \le D_h, \label{constraint:delay}\\
    &q^h_{f} \le r_{\phi_{k},k}, k \in \left[m_f, m_f + d_{f,\phi_{k}}^h - 1 \right],  f=1,\ldots,l_h, \label{constraint:resource1}\\
    &g^h_f \le z_{\phi_{k},k}, k \notin \left[m_f, m_f + d_{f,\phi_{k}}^h - 1 \right],  f=1,\ldots,l_h,  \label{constraint:resource2}\\
    &f \in \mathbb{F}_{\phi_{m_f}}, f = 1, \ldots, l_h, \label{constraint:VNFinstallation}\\
    &e_{\phi_k, \phi_{k+1}}(k) = 1, k \in \left[1, |\boldsymbol{\phi}|\right], \text{ for } \phi_k \ne \phi_{k+1}. \label{constraint:ISL}
\end{align}
\noindent Constraint (\ref{constraint:delay}) guarantees that the end-to-end delay of the placement does not exceed the maximum delay tolerance of SFC $h$. Constraint (\ref{constraint:resource1}) ensures that the satellite executing VNF $f$ at time slot $m_f$ has sufficient computational resources. Constraint (\ref{constraint:resource2}) ensures that every satellite handling the request has enough storage resources to store VNFs' outputs. We note that before the first VNF is executed, only the request event is transferred between satellites, with no intermediate data produced by VNFs; hence, storage resource is considered to be unaffected and the constraint (\ref{constraint:resource2}) is not required for $k \in [1, \max(1,m_1-1)]$. Constraint (\ref{constraint:VNFinstallation}) ensures that only satellites installed with VNF $f$ can execute this VNF. Finally, constraint (\ref{constraint:ISL}) guarantees that data is transferred only via active ISLs. Additionally, we note that if $\phi_k = \phi_{k+1}$, then $e_{\phi_k, \phi_{k+1}}(k) = 1$ always.

\subsection{Multi-Agent Q-Learning Solution} 
\label{sec:learning_model}    
Finding a solution for the DP Eq. (\ref{eq:DP}) subjected to constraints (\ref{constraint:delay})-(\ref{constraint:ISL}) can be computationally intractable due to its recursive nature. Furthermore, this equation assumes knowledge of the probabilities associated with service requests, i.e., $\mu$, $\mu_v^r$, and $\mu_h^s$, which are typically unobtainable in a realistic environment. To tackle this challenge, we reformulate the VNF placement problem as an MAQL model. In this context, each satellite operates as an independent agent with the goal of learning the optimal VNF placement policy for a requested service. The components of the MAQL model are defined as:

\subsubsection{Agent's State}
The agent's state $\mathbf{s}_v$ of a satellite $v \in \mathbb{V}$ in a time slot $t$, describes its available resources and the SFC requests it is currently handling. This state is derived from the system state $\mathbf{x}$, introduced in the previous section for the DP formulation. We note that although only one request arrives at the system per time slot, each satellite might be handling multiple requests at each given moment. An agent's state is defined as follows:
\begin{align}\label{eq:sv}
    \mathbf{s}_v = \{(r_v, z_v), \mathbf{y}_i \; | \; \forall i \in \{1, \ldots, |\mathbb{Y}_v| \}\},
\end{align}
where $r_v$ and $z_v$ are the available computational and storage resources at satellite $v$, respectively, in the considered time slot, $\mathbb{Y}_v$ is the set of requests handled by satellite $v$ at the considered time slot, and $|\mathbb{Y}_v|$ is the number of its elements; $\mathbf{y}_{i} \in \mathbb{Y}_v$ is the $i^{\text{th}}$ handled request from the set of requests handled by satellite $v$ and is defined as:
\begin{align}\label{eq:sat_bufer_req}
    \mathbf{y}_i = \left\{h_i, n_i, f_i, \hat{t}_i \right\},
\end{align}
where $h_i \in \mathbb{H}$ and $n_i \in \mathbb{V}$ is the augmented requested SFC and requester satellite notation for the specific request, respectively; $\hat{t}_i$ represents the timestamp that indicates the initiation time slot of the request. We assume that the requests are stored sequentially in the agent's state, i.e., if $\hat{t}_i < \hat{t}_j$, then $i < j\in [1,|\mathbb{Y}_v|]$. $f_i$ indicates the next VNF in SFC $h_i$ that needs to be executed. When $f_i$ exceeds the total number of VNFs in SFC $h_i$, i.e., $f_i = l_{h_i} + 1$, all the VNFs of the requested SFC $h_i$ have been executed and the final result is to be returned to the requester. An example follows:
\begin{example}
    We consider two SFCs with the following topologies; SFC 1: VNF 2 $\rightarrow$ VNF3 and SFC 2: VNF 2 $\rightarrow$ VNF 1 $\rightarrow$ VNF 3. The execution of each VNF in these SFCs spans $1$ time slot. The following scenario takes place: 
    \begin{itemize}
        \item $t=1$: Satellite $1$ requests for SFC 2, and carries the request over to the next time slot due to inactive ISL.
        \item $t=2$: Satellite $2$ requests for SFC 1 and executes $F_1^1$. Satellite $1$ again carries the request over to the next time slot.
        \item $t=3$: The $(1,3)$- and $(2,3)$-ISLs of the considered satellites are active. Satellite $1$ forwards the request for SFC 2 to satellite $3$, without executing any VNF. Satellite $2$ also forwards the request for SFC 1 and the output of $F_1^1$ to satellite $3$. 
    \end{itemize}
    Therefore, at time slot $t=4$, satellite $3$ is handling two requests $\mathbf{y}_1 = \{2, 1, 1, 1 \}$ and $\mathbf{y}_2 = \{1, 2, 2, 2 \}$ in ascending order of their initiation times. We present Fig. \ref{fig:example3_illustrate} to aid the intuition of this example. $\square$

    \captionsetup[figure]{
  %labelformat=simple,
  %labelsep=colon,
  textfont=normal, % Caption text will not be bold
  %justification=raggedright,
  labelfont={color=black}, % Label will be blue and bold
  skip=10pt % Adjust vertical space if needed
}
    \begin{figure}
	\centering
	%\captionsetup{justification=centering}
	\includegraphics[scale=0.45]{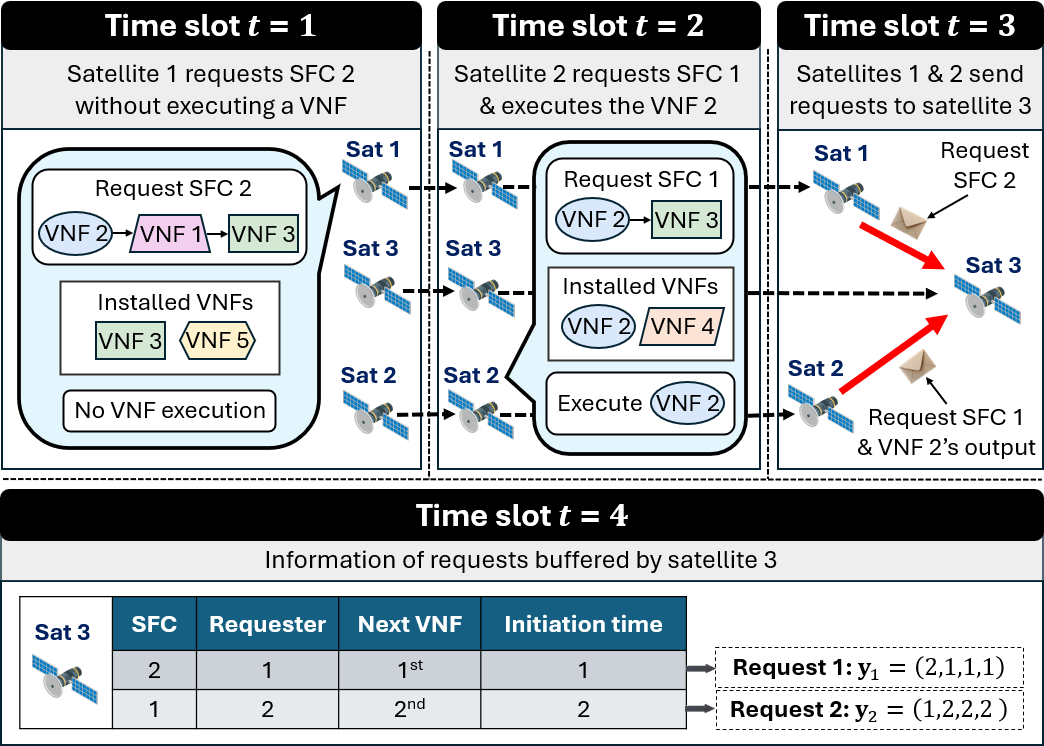}
	\caption{Visualization of the request transfer and satellites' buffered requests in Example 3.}
\label{fig:example3_illustrate}
\end{figure}
    
\end{example}

\subsubsection{Agent's Action} 
Each agent's/satellite's $v \in \mathbb{V}$ VNF placement action at a given time slot is a set of sub-actions, each directed to one of the requests it is currently handling:
\begin{align}\label{eq:action_sequence}
    \mathbf{a}_v = \left\{a_i \; | \; a_i \in \{1, \ldots, V+2 \right\}, \forall i \in \{1, \ldots, |\mathbb{Y}_v|\} \},
\end{align}
where $a_i$ is the sub-action concerning handled request $\mathbf{y}_i$. We remind that the request index $i$ is assigned following ascending order of request initiation time. Based on that, request $i$ is handled before request $i+1$, i.e., decision $a_i$ is taken before $a_{i+1}$. We identify $V+2$ possible sub-actions for each request $\mathbf{y}_i$ and classify them into four categories denoted by $({\sf {A1}}), ({\sf {A2}}), ({\sf {A3}})$, and $({\sf {A4}})$ as follows:
\begin{itemize}
    \item ({\sf {A1}}): $a_i = u \in \mathbb{V} \backslash \left\{v\right\}$; \textit{forward } the request to satellite $u$. This sub-action is valid if the followings hold: (i) the $(v,u)$-ISL is active at the current time slot, and (ii) if $f_i \ge 2$, $z_u \ge g_{f_i}^{h_i}$, i.e., satellite $u$ has sufficient storage resource to receive the request.
    \item ({\sf {A2}}): $a_i = V+1$; \textit{execute} VNF $F^{h_i}_{f_i}$. This sub-action is carried out, provided that the following three conditions are satisfied: (i) not all the VNFs of SFC $h_i$ have been executed, i.e., $f_i \le l_{h_i}$; (ii) the said VNF is pre-installed on satellite $v$, i.e., $F^{h_i}_{f_i} \in \mathbb{F}_v$; (iii) the satellite has sufficient available resources. 
    \item ({\sf {A3}}): $a_i = V+2$; \textit{reject} the request (always valid).
    \item ({\sf {A4}}): $a_i = v$: \textit{stall} and \textit{carry} the request over to the next time slot (always valid).
\end{itemize}
Let $\mathbb{A}_v\left(\mathbf{y}_i\right)$ be the family of valid actions of $v \in \mathbb{V}$ for request $\mathbf{y}_i \in \mathbb{Y}_v$. In this MAQL setup, we determine a VNF placement policy, denoted by $\pi_v\left(\mathbf{y}_i\right) \rightarrow a_i$, as a rule that maps each request $\mathbf{y}_i$ to an action $a_i \in \mathbb{A}_v\left(\mathbf{y}_i\right)$. 
    
\subsubsection{Instant Cost} By performing sub-action $a_i$ for a request $\mathbf{y}_i$, a cost $c_v\left(\mathbf{y_i}, a_i\right)$ is induced on satellite $v$, depending on the the type of sub-action:
\begin{align*}
    c_v(\mathbf{y_i}, a_i) =
    \begin{cases}
        \mathds{1}\left\{z_u \ge g^{h_i}_{f_i} \right\} + C_p \; \mathds{1}\left\{z_u < g^{h_i}_{f_i} \right\},& \text{if } ({\sf A1}),\\
        d^{h_i}_{f_i, v},& \text{if } ({\sf A2}),\\
        C_p,& \text{if } ({\sf A3}),\\
        1,& \text{if } ({\sf A4}),
    \end{cases}
\end{align*}
\noindent where $\mathds{1}\{\cdot \}$ is the indicator function which is equal to $1$ if the enclosed condition is satisfied and $0$ otherwise. In detail, if satellite $v$ forwards the request to satellite $u\in \mathbb{V}$ and $u$ has enough available storage to receive it, then, the induced cost is equal to the transferring time ($1$ time slot). Otherwise, data cannot be received, and a penalty $C_p$ is incurred. In case satellite $v$ executes this part of the handled request, the cost is equal to the VNF's execution time. Rejecting a request costs a penalty equal to $C_p$. Finally, stalling and carrying a request over to the next time slot incurs a cost equal to $1$. Given a satellite's $v$ state $\mathbf{s}_v$, we define the total cost for induced for performing a sequence of sub-actions, i.e., action $\mathbf{a}_v$ as:
\begin{align}
    C_v\left(\mathbf{s}_v, \mathbf{a}_v\right) = \sum\limits_{i=1}^{|\mathbb{Y}_v|} \gamma^{\hat{t}_i+1} c_v\left(\mathbf{y}_i, a_i \right).
\end{align}

\subsubsection{Learning Mechanism \& Optimal Action Estimation} In the following, we augment the agent's state, handled request and agent's action notation to accommodate time slot locality, as $\mathbf{s}(t)$, $\mathbf{y}(t)$ and $\mathbf{a}(t)$. We additionally drop the satellite, $v$, and sequence, $i$, subscripts from them, to avoid overloading the equations. Given the above, in the MAQL environment we are defining, each satellite $v \in \mathbb{V}$ aims at learning the optimal VNF placement policy that minimizes the following discounted long-term cost:
\begin{align}
    &\sum_{t=1}^{\infty}  \mathbb{E}\left[C_v\left(\mathbf{s}\left(t\right), \mathbf{a}\left(t\right) \right) \right] = \sum_{ \mathbf{y}\left(t\right) \in \mathbb{Y}_v} \Psi(\mathbf{y}(t), a(t)), \label{eq:longterm_cost_maql}
\end{align}
where $\Psi(\mathbf{y}(t), a(t)) = \sum_{t=1}^{\infty}  \mathbb{E}_{\pi_v}\left[\gamma^{\hat{t}+1}c_v\left(\mathbf{y}\left(t\right), a\left(t\right) \right) \right]$, 
$a(t)$ being the sub-action for request $\mathbf{y}(t)\in \mathbb{Y}_v$ dictated by policy $\pi_v$. Approximating the optimal policy is equivalent to approximating an action for every request in each time slot that minimizes this discounted long-term cost. 
    
In our MAQL framework, the \textit{Q-table} contains the approximations of the long-term sub-costs $\Psi(\mathbf{y}(t), a(t))$. Let $\mathbb{Q}_v$ be the Q-table of satellite $v$ that stores all the learned Q-values. We remind the reader that the network topology varies deterministically and periodically with a period $T$. Therefore, at a time slot $t$, the network topology and its upcoming changes are the same as those at a time slot $\zeta$, where $1 \le \zeta \le T$. This time slot $\zeta$ then can be computed wrt. $t$ as $\zeta = t \; \mathds{1}\left\{t < T\right\} + \text{mod}\left(t, T\right)\mathds{1}\left\{t \ge T\right\}$,
where $\text{mod}\left(t, T\right)$ returns the remainder of $t \div T$. To this end, we denote by  $Q_v^{\left(\zeta\right)}\left(\mathbf{y}\left(t\right), a\left(t\right)\right)$ the Q-value estimation of cost $\Psi(\mathbf{y}(t), a(t))$. For presentation convenience %and simplicity, for the rest of this section, 
we fix the time slot $t$ thus hereafter the $t$ notation is dropped.

In this setting, each satellite's environment is non-stationary as it depends on the policies employed by other satellites which continually evolve, meaning that system convergence is not guaranteed. To address this concern, we promote collaboration among the agents in the form of knowledge sharing. When satellite $v \in \mathbb{V}$ forwards a request $\mathbf{y}\in \mathbb{Y}_v$ to satellite $u \in \mathbb{V} \setminus \{v\}$, $u$ shares its learned Q-values as follows:
\begin{itemize}
    \item If satellite $u$ has handled the request $\mathbf{y}$ before, i.e., the row $Q_u^{\left(\zeta\right)}\left(\mathbf{y}, a \right)$ is populated on $\mathbb{Q}_u$, then the value $\underset{a}{\min}~Q_u^{\left(\zeta\right)}\left(\mathbf{y}, a \right)$ is shared with satellite $v$. %This shared value corresponds to the minimum discounted cost that satellite $u$ can achieve for request $\mathbf{y}\in \mathbb{Y}_v$.
    \item Otherwise, a predefined initial Q-value $Q_{init}$ is shared.
\end{itemize}
Let $\mathbf{Q}_v^{\left(\zeta\right)}\left(\mathbf{s}_v, \mathbf{a}\right) = \{Q_v^{\left(\zeta\right)}\left(\mathbf{y}, a\right) |~ \forall \mathbf{y} \in \mathbb{Y}_v\}$. Having acquired shared Q-values from other satellites, satellite $v$ updates each component $Q_v^{\left(\zeta\right)}\left(\mathbf{y}, a\right)$ if $a \ne V+2$ as follows:
\begin{align}
    \label{eq:maql_Qupdate}
    Q_v^{\left(\zeta\right)}\left(\mathbf{y}, a\right)  \leftarrow \left(1-\lambda\right)Q_v^{\left(\zeta \right)}\left(\mathbf{y}, a\right)  + \lambda(c\left(\mathbf{y}, a\right) +  \delta \hat{Q} ).
\end{align}
$\hat{Q} = \underset{a' \in \mathbb{A}_w\left(\mathbf{y}'\right)}{\min}~Q_v^{\left(\zeta\right)}\left(\mathbf{y}', a' \right),$ where $w=v$ if $a \in \{v, V+1\}$; $w=u$ if $a \in \{1,\ldots,V \} \backslash \{v\}$. $\lambda$, $\delta \in [0, 1]$ are the learning rate and the discount factor, respectively; $\mathbf{y}'$ is the next state of $\mathbf{y} = \{h, n, f, \hat{t}\}$, where: (i) $\mathbf{y}' = \mathbf{y}$ if $a \ne V+1, V+2$, (ii) $\mathbf{y}' = \{h, n, f+1, \hat{t}\}$ if $a = V+1$. If $a = V+2$, the request is rejected and the penalty cost $C_p$ is yielded: $Q_v^{\left(\zeta\right)}\left(\mathbf{y}, V+2\right)=C_p$. Then, satellite $v$ approximates its optimal action $a^*$ for a handled request $\mathbf{y}\in \mathbb{Y}_v$ by:
\begin{align}
    \label{eq:best_action}
    a^* = \underset{a \in \mathbb{A}_v\left(\mathbf{y}\right)}{\argmin}~Q_v^{\left(\zeta\right)}\left(\mathbf{y}, a\right).
\end{align}
A formal description of the MAQL-based VNF placement method is provided in Algorithm \ref{algo:MAQL}. %We remind that $\mathbb{Q}_v$ is the Q-table of satellite $v$, and $Q_v^{(\zeta)}(\mathbf{y}_i, a_i)$ a Q-value in it.}
\begin{algorithm} 
	\caption{Cooperative VNF Placement on Satellite $v$} \label{algo:MAQL}
	\begin{algorithmic}[1] 
		%\Procedure{MyProcedure}{}
        \State \textbf{Initialize} $t \leftarrow 1$, %$\mathbb{Y}_v \leftarrow \emptyset$ 
        \State $e_{v,u}(1) \leftarrow 1$ if $(v,u)$-ISL is active, $0$ if not, $\forall u \in \mathbb{V} \backslash \{v \}$
        \While{$\mathbb{Y}_v \ne \emptyset$}
            \State \textit{Select} $\mathbf{y}_i=(h_i, n_i, f_i, \hat{t}_i) \in \mathbb{Y}_v$ %following increasing $\hat{t}$.
            \State $\mathbb{A}_v(\mathbf{y}_i) \leftarrow \{u \ne v \; | \; e_{v,u}(t) = 1 \} \cup \{v \} \cup \{V+2 \}$
            \If{$F^{h_i}_{f_i} \in \mathbb{F}_v$}
            \State $\mathbb{A}_v(\mathbf{y}_i) \leftarrow \mathbb{A}_v(\mathbf{y}_i) \cup \{V+1 \}$
            \EndIf
            \If{\textit{training}}
            \State \textit{Sample} action $a_i \in \mathbb{A}_v(\mathbf{y}_i)$ uniformly
            \Else \State\textit{Perform} $a^*$, Eq. \eqref{eq:best_action}
            \EndIf
            %\State Flag $\mathbf{y}$ as it has received an action.
        \EndWhile
        \For{$u \in \mathbb{V} \backslash \{v\}$ \textbf{such that} $e_{v,u}(t)=1$}
        %\State \textbf{For every satellite} $u \in \mathbb{V} \backslash \{v\}$ \textbf{such that} $e_{vu}(t)=1$\textbf{, do the following:}
            \State \textit{Forward} requests to $u$, $\forall \mathbf{y}_i$ for which $a_i=u$
            \For{$\mathbf{y}_i \in \mathbb{Y}_u$ transferred by $u$ with $i = 1, 2, \ldots$}
                \State $\mathbb{Y}_v \leftarrow \mathbb{Y}_v \cup \{\mathbf{y}_i \}$ if sufficient storage
            \EndFor
            %\State $\mathbb{Y}_v \leftarrow \mathbb{Y}_v \cup \{\mathbf{y}_i |~ \forall \mathbf{y}_i \text{ transferred by } u\}$ %\Comment{(Receive requests from $u$)}
            \State \textit{Receive} Q-value $Q_u^{(\zeta)}(\mathbf{y}_i, a_i)$ from $u$,  $\mathbf{y}_i$: $a_i = u$
            \State $\mathbb{Y}_v \leftarrow \mathbb{Y}_v \backslash \{\mathbf{y}_i \in \mathbb{Y}_v \; | \; \mathbf{y}_i \text{ has been assigned $a_i = u$}\}$
            \State \textit{Update} Q-table $\mathbb{Q}_v$ of satellite $v$ using Eq. (\ref{eq:maql_Qupdate})
        \EndFor
        \State $t \leftarrow t + 1$
        \State $e_{v,u}(t) \leftarrow 1$ if $(v,u)$-ISL is active, $0$ if not, $\forall u \in \mathbb{V} \backslash \{v\}$
        \State \textbf{go to} line 3
		%\EndProcedure
	\end{algorithmic}
\end{algorithm}

Mapping the policies $\pi_v, \forall v\in \mathbb{V}$, resulted from the convergence of the MAQL algorithm, back to $\mathbf{p}$, allows us to define $\Pi: \mathbf{x} \rightarrow \mathbf{p}$ as the VNF placement policy that minimizes end-to-end delay for the requested service.

%\textcolor{blue}{We note that the initial performance of the proposed MAQL framework could be low at the beginning of the training process. For means of practical implementation, a well-established VNF placement scheme can be temporarily adopted. Meanwhile, the MAQL model is trained in the background with collected data. Certain randomness in selecting action needs to be introduced to the employed policy as an exploration activity to aid the convergence of the MAQL model. Once the MAQL model's performance reaches a desirable level, it alters the temporary scheme.}

 \setlength{\parindent}{0pt}{\textbf{Remark 1. (MAQL Framework's Features Supporting Large-Scale LEO Satellite Systems)}: \textit{In mega-constellation LEO satellite networks, the decentralized architecture of the MAQL framework facilitates scalability and reduces the need for extensive global coordination. Furthermore, parameter sharing mechanism enhances both scalability and learning efficiency within large networks.}}

\comment{
\begin{remark}
In large-scale systems, the decentralized nature of the MAQL framework promotes scalability and minimizes the need for centralized global coordination. Additionally, knowledge sharing among satellites further enhances scalability and learning efficiency in large networks. Moreover, learning parameters are transferred along with service requests. This helps mitigate the often significant overhead of large-scale systems.
\end{remark}
}

\section{VNF Caching in LSNs}
\label{sec:VNF_caching}    
In the previous sections, we assumed that each satellite comes pre-installed with a subset of VNFs. Based on this, VNF placement is carried out. In this section, we aim to determine the optimal subset of VNFs to be cached/pre-installed on each satellite, thereby maximizing the successful request serving rate.  

\subsection{Objective Function Formulation}
%We denote by $\Theta_v$ the caching capacity at satellite $v\in \mathbb{V}$, i.e., the maximum number of VNFs that can pre-installed. 
%We denote by $F_i$ the VNF of type $i$. Due to the notation complexity, we re-define $\mathbb{F}_v$ as the set of VNF types cached at satellite $v$. 
We recall that $|\mathbb{F}_v|$ is the given, fixed caching capacity of satellite $v$. A system-wide caching strategy is  given as:
\begin{align}
    \boldsymbol\theta = \left\{\theta_i | \theta_i \in \mathbb{F}, \forall i \in \left[1,  \sum_{u=1}^V |\mathbb{F}_u| \right] \right\},
\end{align}
where $\mathbb{F}_v = \left\{\theta_i \; | \; \forall i\in \left[\sum_{u=1}^{v-1} |\mathbb{F}_u|+1, \sum_{u=1}^v |\mathbb{F}_u|\right] \right\}$ contains the VNFs cached by satellite $v \ge 2$, and $\mathbb{F}_1 = \{\theta_i \; | \; \forall i\in \left[1, |\mathbb{F}_1|\right] \}$ are those cached by satellite 1. The objective of the VNF caching task is to maximize the average number of requests served in a time slot, i.e., the request serving rate. Thus, we aim at maximizing the objective function $\mathbb{E}\left[ M_{\Pi}(\boldsymbol\theta)\right]$ where $\mathbb{E}\left[\cdot \right]$ is the expectation operator; $M_{\Pi}(\boldsymbol\theta)$ is associated with the VNF placement policy $\Pi$ and caching strategy $\boldsymbol\theta$, as:
\begin{align}
    \label{eq:caching_objfunction}
    M_{\Pi}(\boldsymbol\theta) = \frac{1}{\Omega}\sum_{t = 1}^{\Omega} \mathds{1}\left\{ \mathcal{C}\left(\mathbf{x}(t), \Pi(\mathbf{x}(t) \; | \; \boldsymbol\theta)) \right) < C_p \right\},
\end{align}
where $\Omega$ is a predefined time horizon for calculating the serving rate and $\Pi(\mathbf{x}(t) \; | \; \boldsymbol\theta)$ is the VNF placement at time slot $t$, following policy $\Pi$, given the VNF caching strategy $\boldsymbol\theta$. Our objective is then to calculate the optimal caching strategy that maximizes the expected request serving rate:
\begin{align}
\label{eq:caching-objective}
    \boldsymbol\theta^* = \underset{\boldsymbol\theta}{\text{arg} \,\max}~ \mathbb{E}\left[M_{\Pi}(\boldsymbol\theta) \right].
\end{align}

\subsection{Bayesian Optimization Solution}
\label{sec:BOA}
Bayesian Optimization (BO) is a powerful iterative method for optimizing black-box functions \cite{garnett_2023}. This subsection outlines the key components and the BO-based algorithm for solving Eq. \eqref{eq:caching-objective}.
    
\subsubsection{Search Space} 
We define our solution search space as the family $\boldsymbol{\Theta}$ of caching strategies $\boldsymbol\theta$  that satisfy the following conditions: i) there exists at least one SFC $h\in \mathbb{H}$ such that $\mathbb{S}_h \subseteq \boldsymbol\theta$ and ii) a VNF is not cached more than once in each satellite.\footnote{In executing multiple VNFs, a satellite can access simultaneously access all installed VNFs. Hence, it is sufficient to cache a VNF once.} The first condition ensures that at least requests for one type of SFCs can be served, and the second guarantees that no caching storage is wasted. 

\subsubsection{Surrogate Model} A probabilistic, surrogate model is selected and trained to predict the distributions of the objective function's outcomes, defined in Eq. (\ref{eq:caching_objfunction}), for various inputs $\boldsymbol\theta \in \boldsymbol{\Theta}$. Let $\dot{\mu}(\boldsymbol\theta)$ and $\dot{\sigma}(\boldsymbol\theta)$ be the predicted mean and standard deviation for $\boldsymbol\theta$, respectively. Let also the radial basis function $\mathcal{K}(\boldsymbol\theta, \boldsymbol\theta') = \exp\left(\frac{1}{2\beta^2} ||\boldsymbol\theta - \boldsymbol\theta' || \right),$ serve as the kernel function, which is responsible for measuring how close the VNF caching strategy is to the training data,
where $\beta$ is a predefined length scale parameter and $\boldsymbol\theta' \neq \boldsymbol\theta$. We select the Gaussian Process (GP) as our surrogate model based on the following considerations.

First, $M_{\Pi}(\boldsymbol\theta)$ defined in Eq. (\ref{eq:caching_objfunction}) is a summation of the indicator functions, the outputs of which are random quantities.
%$M_{\Pi}(\boldsymbol\theta)$ does not guarantee a normal distribution, the outputs of the indicator function in Eq. (\ref{eq:caching_objfunction}) are random variables, leading $M_{\Pi}(\boldsymbol\theta)$ to be the sum of multiple random variables.
Then, when $\Omega$ is sufficiently large and the outputs of the indicator function are weakly dependent, the Central Limit Theorem \cite{allofstatistics} can be applied. This theorem suggests that, under these conditions, the distribution of $M_{\Pi}(\boldsymbol\theta)$ will approximate a normal distribution. %although the normality of the distribution of the objective function $M_{\Pi}(\boldsymbol\theta)$ is not guaranteed, the outputs of the indicator function in Eq. (\ref{eq:caching_objfunction}) are random, thus $M_{\Pi}(\boldsymbol\theta)$ is the sum of multiple random variables. Therefore, in cases where $\Omega$ is sufficiently large and the outputs of the indicator function are weakly dependent random variables, the Central Limit Theorem \cite{allofstatistics} can be applied, making the distribution of $M_{\Pi}(\boldsymbol\theta)$ a normal distribution.

Second, GP modeling offers computational tractability. Let $\tilde{\boldsymbol\theta} \in \boldsymbol{\Theta}$ be an input strategy for which $M_{\Pi}(\tilde{\boldsymbol\theta})$ has been evaluated. We call $\tilde{\boldsymbol\theta}$ a training data point, and $\tilde{\boldsymbol{\Theta}} = \{\tilde{\boldsymbol\theta}_i \; | \; \forall i \in [1, |\tilde{\boldsymbol{\Theta}}|\,] \}$ the training dataset. Under the GP modeling, we have the \textit{prior distribution} $M_{\Pi}(\tilde{\boldsymbol\theta}) \sim \mathcal{N}\left(\dot\mu(\tilde{\boldsymbol\theta}), \mathcal{K}(\tilde{\boldsymbol{\Theta}}, \tilde{\boldsymbol{\Theta}}) \right)$ where $\mathcal{K}(\tilde{\boldsymbol{\Theta}}, \tilde{\boldsymbol{\Theta}})$ is a matrix consisting of elements $\mathcal{K}(\tilde{\boldsymbol\theta}, \tilde{\boldsymbol\theta}')$, $\forall \tilde{\boldsymbol\theta} \neq \tilde{\boldsymbol\theta}' \in \tilde{\boldsymbol{\Theta}}$. For a family of non-evaluated strategies $\mathbb\Delta \subseteq \boldsymbol{\Theta}$, the joint distribution between $M_{\Pi}(\tilde{\boldsymbol{\Theta}})$ and $M_{\Pi}(\mathbb\Delta)$ can be expressed as:
\begin{align*}
    \binom{M_{\Pi}(\tilde{\boldsymbol{\Theta}})}{M_{\Pi}(\mathbb\Delta)} \sim \mathcal{N}\left( \binom{\dot\mu\left(\tilde{\boldsymbol{\Theta}} \right)}{\dot\mu\left(\mathbb\Delta \right)},  \binom{\mathcal{K}(\tilde{\boldsymbol{\Theta}}, \tilde{\boldsymbol{\Theta}}) ~ \mathcal{K}(\tilde{\boldsymbol{\Theta}}, \mathbb\Delta)}{\mathcal{K}(\mathbb\Delta, \tilde{\boldsymbol{\Theta}}) ~ \mathcal{K}(\mathbb\Delta, \mathbb\Delta)} \right)
\end{align*}
that provides analytically tractable expressions for the marginal and conditional distributions. The \textit{posterior distribution}, i.e., the predicted distribution of non-evaluated strategies $\mathbb\Delta$, given the training dataset $\tilde{\boldsymbol{\Theta}}$, follows a Gaussian distribution, $\mathcal{N}\left(\dot\mu(\mathbb\Delta), \dot\sigma(\mathbb\Delta) \right)$, where:
\begin{align}
    &\dot\mu(\boldsymbol\theta) = \dot\mu(\tilde{\boldsymbol{\Theta}}) + \mathcal{K}(\boldsymbol\theta, \tilde{\boldsymbol{\Theta}})^\intercal \mathcal{K}(\tilde{\boldsymbol{\Theta}}, \tilde{\boldsymbol{\Theta}})^{-1}(M_{\Pi}(\tilde{\boldsymbol{\Theta}}) - \dot\mu(\tilde{\boldsymbol{\Theta}})) \label{eq:mean_posterior},\\
    &\mathcal{K}(\tilde{\boldsymbol\theta}, \boldsymbol\theta) = \mathcal{K}(\tilde{\boldsymbol\theta}, \tilde{\boldsymbol\theta}) + \mathcal{K}(\tilde{\boldsymbol\theta}, \tilde{\boldsymbol{\Theta}})^ \intercal \mathcal{K}(\tilde{\boldsymbol{\Theta}}, \tilde{\boldsymbol{\Theta}})^{-1}\mathcal{K}(\boldsymbol\theta, \tilde{\boldsymbol{\Theta}})^ \intercal  \label{eq:k_posterior},
\end{align}
$\boldsymbol\theta \in \mathbb\Delta$, and $\boldsymbol\theta \notin \tilde{\boldsymbol{\Theta}}$ being a non-evaluated caching strategy and $\tilde{\boldsymbol\theta}\in \tilde{\boldsymbol{\Theta}}$ being a strategy from the training dataset. The obtained posterior distribution is used to predict the mean and variance of $M_{\Pi}(\boldsymbol\theta)$. The current posterior distribution is then considered as prior in the following iteration.

\subsubsection{Acquisition Function} The role of the acquisition function in the BO is to guide the search for the optimal solution in the search space \cite{garnett_2023}. To come up with the most efficient way of doing so, we implemented and evaluated the following four acquisition functions: 

\textit{a) Probability of Improvement (PI):} this function provides an exploration-exploitation balance metric for guiding the search process by estimating the likelihood that a new sample will surpass the current best observation, $M_{\sf max} = \underset{\tilde{\boldsymbol\theta} \in \tilde{\boldsymbol{\Theta}}} {\max}~ M_{\Pi}(\tilde{\boldsymbol\theta})$. %this function quantifies the probability that the objective function's output will improve compared to the current best value, $M_{\sf max} = \underset{\tilde{\boldsymbol\theta} \in \tilde{\boldsymbol{\Theta}}} {\max} M_{\Pi}(\tilde{\boldsymbol\theta})$, for a candidate input strategy $\boldsymbol\theta$.
The PI function is defined as:
\begin{align}
    I(\boldsymbol\theta) =1 - \Phi\left(\frac{M_{\sf max} - \dot\mu(\boldsymbol\theta)}{\dot\sigma({\boldsymbol\theta})} \right) \label{eq:PoI_func}
\end{align}
where $\Phi$ is the Cumulative Distribution Function (CDF) of the standard normal distribution.

\textit{b) Expected Improvement (EI):} this function measures the expected improvement over $M_{\sf max}$, when evaluating a candidate input strategy $\boldsymbol\theta$, and is defined as:
\begin{align}
    \label{eq:EI_func}
    I(\boldsymbol\theta) = \mathbb{E}\left[\max(M_{\Pi}(\boldsymbol\theta) - M_{\sf max}, 0) \right].
\end{align}

\textit{c) Upper Confidence Bound (UCB):} This function aims to balance exploration and exploitation by considering the uncertainty associated with the objective function's evaluations and is given by:
\begin{align}\label{eq:UCB_func}
    I(\boldsymbol\theta) = \dot{\mu}(\boldsymbol\theta) + \xi \; \dot{\sigma}(\boldsymbol\theta),
\end{align}
where $\xi$ is a predefined parameter that controls the trade-off between exploration and exploitation.

\textit{d) Lower Confidence Bound (LCB):} This function aims to maximize the lower estimate of the objective function's evaluations and is defined as:
\begin{align}
    \label{eq:LCB_func}
    I(\boldsymbol\theta) = \dot{\mu}(\boldsymbol\theta) - \xi \; \dot{\sigma}(\boldsymbol\theta),
\end{align}
LCB tends to exploit the known regions of the search space more aggressively.
    
\subsubsection{The BO Algorithm} We initiate the process by constructing the training set $\tilde{\boldsymbol{\Theta}}$; Depending on the size $|\tilde{\boldsymbol{\Theta}}|$ we are opting for, we repeat a random selection of candidate caching strategies from the search space, $\boldsymbol\theta \in \boldsymbol{\Theta}$ for evaluation; $M_{\Pi}(\boldsymbol\theta)$ is calculated by caching the VNFs on satellites according to $\boldsymbol\theta$, letting the system run for $\Omega$ time slots following the VNF placement policy $\Pi$, and counting the number of successfully served requests. Naturally, this makes the evaluation step a time-consuming process. Therefore, during the BO runtime, we only evaluate caching strategies $\boldsymbol\theta$ that have the potential to be the optimal solution, $\boldsymbol\theta_{\sf max}$, guided by the acquisition function:
\begin{align}\label{eq:nextpoint_acquif}
    \boldsymbol\theta_{\sf max} = \underset{\boldsymbol\theta \in \boldsymbol{\Theta}}{\text{arg} \,\max}~I(\boldsymbol\theta),
\end{align}
We then evaluate the objective function $M_{\Pi}(\boldsymbol\theta_{\sf max})$ and update the surrogate model; the prior distribution of $M_{\Pi}(\boldsymbol\theta)$ is updated to a posterior Multi-variate Gaussian Distribution, with a mean value and a kernel matrix given by Eq. (\ref{eq:mean_posterior}) and (\ref{eq:k_posterior}), where $\tilde{\boldsymbol{\Theta}}$ now is the training dataset augmented by $\boldsymbol\theta_{\sf max}$, the last evaluated input strategy. 

After running the above steps for a sufficient number of time slots and candidate optimal solutions $\boldsymbol\theta_{\sf max}$, the optimal VNF caching strategy $\boldsymbol\theta^*$ is calculated as:
\begin{align}
    \label{eq:return}
    \boldsymbol\theta^* = \underset{\boldsymbol\theta \in \tilde{\boldsymbol{\Theta}}}{\text{arg} \,\max}~M_{\Pi}(\boldsymbol\theta).
\end{align}

\noindent In practice, each satellite $v \in \mathbb{V}$ obtains the VNF caching data from the controller on the ground and installs the VNFs dictated by $\boldsymbol\theta^*$. We envision this step as a one-time pre-online task. However, the learning process can continue to operate in the background, during runtime, seeking potential improvements in the VNF caching strategies. Each newly discovered and superior solution can be seamlessly applied, leading to an iterative refinement process. We concisely and formally present the discussed BO-based VNF caching method in Algorithm \ref{algo:BO}.

\begin{algorithm} [H]
	\caption{BO-based VNF Caching} \label{algo:BO}
	\begin{algorithmic}[1] 
		%\Procedure{MyProcedure}{}
        \State \textit{Select} $\dot\mu(\cdot)$ and $\mathcal{K}(\cdot, \mathbf{\cdot})$
        \Comment{(Prior distribution)}
        \State \textbf{until} sufficient $|\tilde{\boldsymbol{\Theta}}|$ \textbf{do} 
        \State ~~~ \textit{Sample} $\boldsymbol\theta \in \boldsymbol{\Theta}$, randomly \& uniformly
        \State ~~~ \textit{Evaluate} $M_{\Pi}(\boldsymbol\theta)$
        \State ~~~ $\tilde{\boldsymbol{\Theta}} \leftarrow \tilde{\boldsymbol{\Theta}} \cup \{\boldsymbol\theta \}$ \Comment{(Construct initial training set)}
        \State \textbf{until} large enough $|\tilde{\boldsymbol{\Theta}}|$ \textbf{do}
        \State ~~~  \textit{Calculate} $\boldsymbol\theta_{\max}$ using Eq. (\ref{eq:nextpoint_acquif})
        \State ~~~  \textit{Evaluate} $M_{\Pi}(\boldsymbol\theta_{\max})$
        \State ~~~ $\tilde{\boldsymbol{\Theta}} \leftarrow \tilde{\boldsymbol{\Theta}} \cup \{\boldsymbol\theta_{\sf max} \}$
        \State ~~~ \textit{Update} $\dot\mu(\boldsymbol\theta)$ and $\mathcal{K}(\tilde{\boldsymbol\theta}, \boldsymbol\theta)$ using Eqs. (\ref{eq:mean_posterior}) and 
        (\ref{eq:k_posterior})
        \State ~~~
        \Comment{(Posterior distribution)}
        \State \textbf{return} $\boldsymbol\theta^*$ using Eq. \eqref{eq:return}
		%\EndProcedure
	\end{algorithmic}
\end{algorithm}

In addition, we note that the proposed VNF placement and the caching frameworks are dependent on each other. The optimal solution is obtained by executing the two processes of caching and placement in an iterative fashion. The BO-based algorithm presented in this subsection serves as an outer layer that provides the caching strategy that needs to be evaluated. The MAQL-based VNF placement policy, which serves as an inner iteration, returns the optimal serving rate after convergence. As we iterate, we obtain an optimal caching strategy and a converged MAQL model associated with this strategy.
 
\section{Numerical Results} 
\label{sec:experiments}
In this section, we assess the performance of the introduced VNF placement and caching algorithms. In this framework, we utilize the concept of time slots to discretize the transition of state $\mathbf{s}_v$ (defined in Eq. (\ref{eq:sv})) and the motion of every satellite $v$ where the latter factor leads to the intermittency of ISLs. For the simulations, the default values of the common parameters are given in Table \ref{tab:defaultval}. Variations in these or other parameters will be noted in each simulation. The details of the considered SFCs are given in Table \ref{tab:SFC_set1} with $D_h=15$ time slots, $\forall h$. The required computational and storage resources are given as follows:
\begin{itemize}
    \item SFC 1: $g_f^1=3,1$ and $q_f^1=5,4$ for $f=1,2$.
    \item SFC 2: $g_f^2=1,3,1$ and $q_f^2=3,4,4$ for $f=1,2,3$.
    \item SFC 3: $g_f^3=1,2$ and $q_f^3=3,3$ for $f=1,2$.
    \item SFC $h\in \{4,\ldots,14\}$: $g_f^h=1$ and $q_f^h=2$, $\forall f$.
\end{itemize}
%To better emphasize each evaluation aspect, we will define logical satellite subgroups where needed throughout the experimentation. Satellites within the same subgroup are assumed to be launched together and share identical speed and orbit characteristics.
\begin{table}[b]
    \caption{Experimental parameter configuration.}
    \label{tab:defaultval}
    \centering
    \begin{tabular}{ |c|c| } 
        \hline
        \textbf{Parameter} & \textbf{Value} \\
        \hline
        $d^h_{f,v}$ & 1 time slot \\ 
        \hline
        $\mu$, $\mu^r_v$, $\mu^s_h$ & 0.9, $1/V$, $1/H$ \\
        \hline
        $\tau_{v,u}$ & 1 time slot \\
        \hline
        $C_p$, $\gamma$, $\delta$ & 100, 0.6, 0.6\\
        \hline
    \end{tabular}
\end{table}
\begin{table}
\caption{SFC index ($h$) and topology ($\mathbf{\mathbb{S}_h}$).}
\label{tab:SFC_set1}
\centering
\begin{tabular}{cc}
\begin{tabular}{|c|c|}
    \hline  
    \textbf{Index} & \textbf{Topology}   \\
    \hline 
    1 & $2 \rightarrow 3$ \\
    \hline 
    2 & $2 \rightarrow 1 \rightarrow 3$ \\
    \hline 
    3 & $1 \rightarrow 3$ \\
    \hline 
    4 & $3 \rightarrow 4$ \\
    \hline 
    5 & $2 \rightarrow 3 \rightarrow 5$ \\
    \hline 
    6 & $1 \rightarrow 4 \rightarrow 6$ \\
    \hline 
    7 & $5 \rightarrow 6$ \\
    \hline
\end{tabular} &  % starting rightmost sub table
% table 2
\begin{tabular}{|c|c|}
    \hline  
    \textbf{Index} & \textbf{Topology} \\
    \hline 
    8 & $7 \rightarrow 8 \rightarrow 9$ \\
    \hline 
    9 & $1 \rightarrow 10$ \\
    \hline 
    10 & $4 \rightarrow 8$ \\
    \hline 
    11 & $6 \rightarrow 10$ \\
    \hline 
    12 & $3 \rightarrow 4$ \\
    \hline 
    13 & $7 \rightarrow 8$ \\
    \hline 
    14 & $1 \rightarrow 2$ \\
    \hline
\end{tabular} \\
\end{tabular}
\end{table}

\subsection{Cooperative MAQL-based VNF Placement}\label{subsec:simu_maql}

\captionsetup[figure]{
  %labelformat=simple,
  %labelsep=colon,
  textfont=normal, % Caption text will not be bold
  %justification=raggedright,
  labelfont={color=black}, % Label will be blue and bold
  skip=10pt % Adjust vertical space if needed
}
\begin{figure*}
    \centering
    %\captionsetup[subfigure]{justification=centering}
    %\captionsetup{justification=centering}
    \begin{subfigure}[b]{0.343\textwidth}
        \centering
        %\captionsetup{justification=centering}
		\includegraphics[width=1\textwidth]{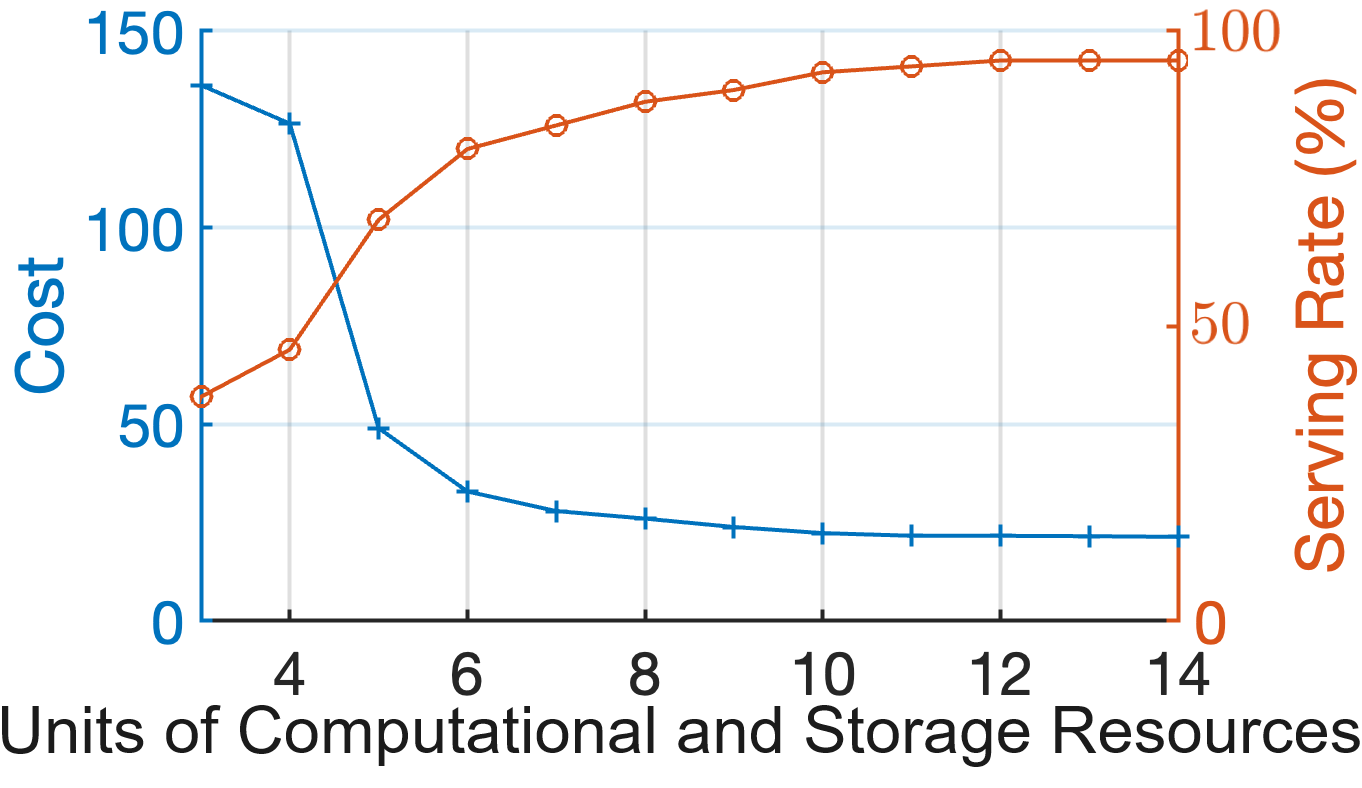}
		%\caption{Single-request context.}
        \caption{Optimal DP cost and serving rate vs resources.}
        \label{fig:dp_setup1}
	\end{subfigure}
    ~% <-- Don't remove these comment to keep horizontal figures
    \begin{subfigure}[b]{0.31\textwidth}
        \centering
	    \includegraphics[width=1\textwidth]{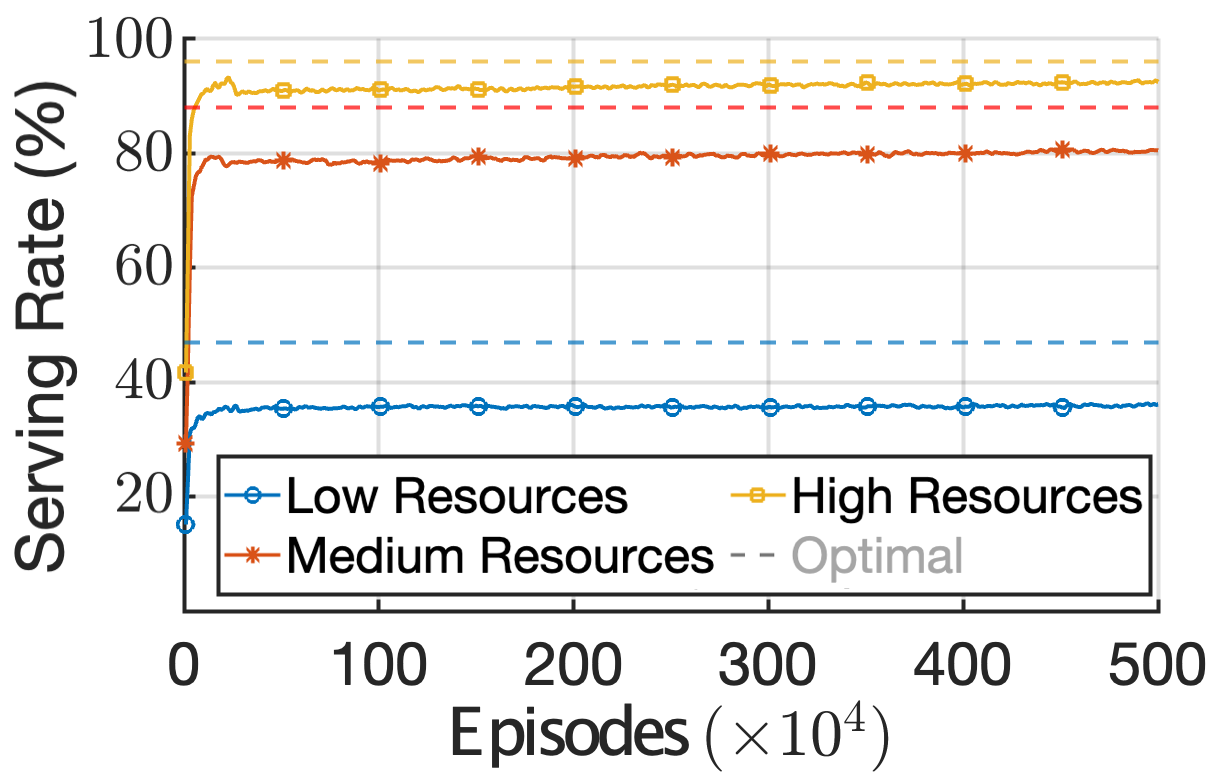}
        %\caption{General context.}
        \caption{Serving rate per training episode vs resources.}
        \label{fig:sr_setup1}
	\end{subfigure}
    \begin{subfigure}[b]{0.32\textwidth}
        \centering
        %\captionsetup{justification=centering}
		\includegraphics[width=1\textwidth]{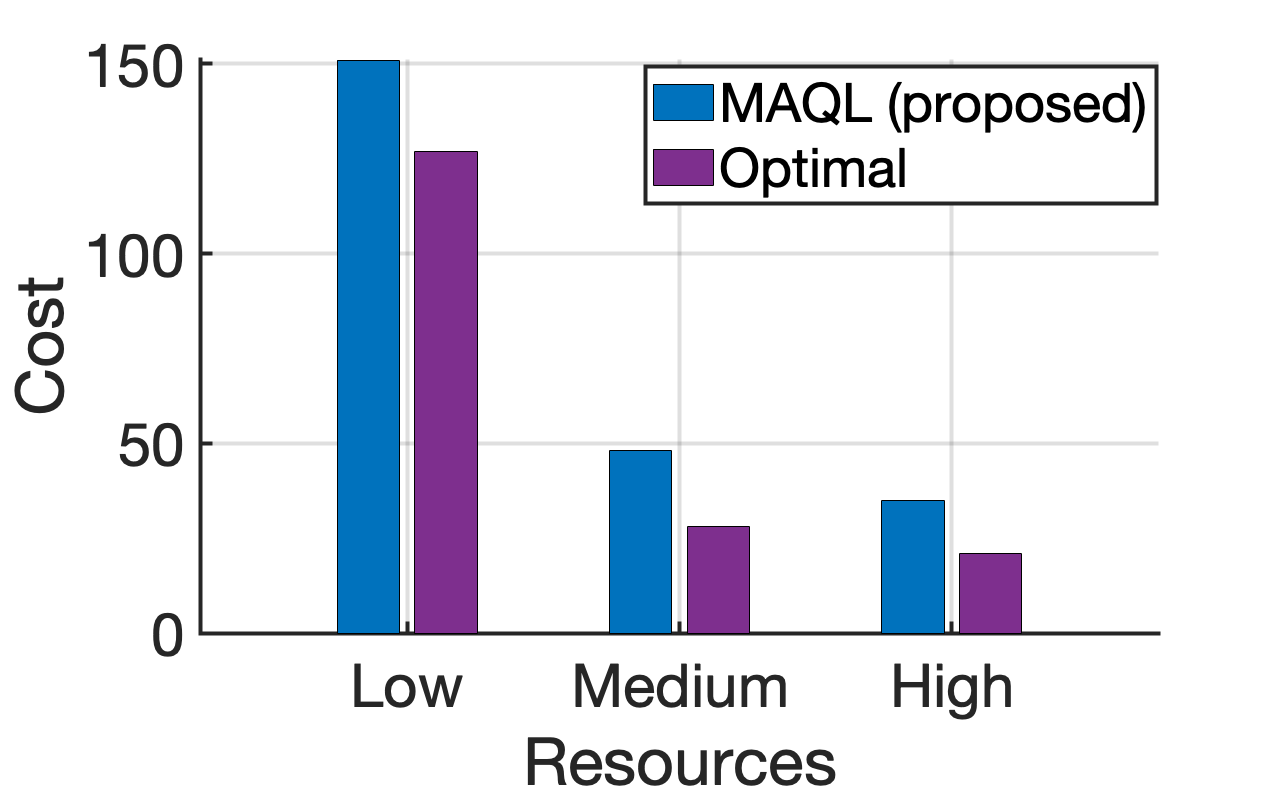}
        %\caption{Impact of $\delta$.}
        \caption{System cost comparison with optimal DP solution.}
		\label{fig:cost_setup1}
    \end{subfigure}
    \caption{Evaluation of the proposed MAQL-based scheme in comparison with the optimal DP solution.}
    \label{fig:setup1}
\end{figure*}

\begin{figure*}
    \centering
    %\captionsetup[subfigure]{justification=centering}
    %\captionsetup{justification=centering}
	\begin{subfigure}[b]{0.325\textwidth}
        \centering
        %\captionsetup{justification=centering}
		\includegraphics[width=1\textwidth]{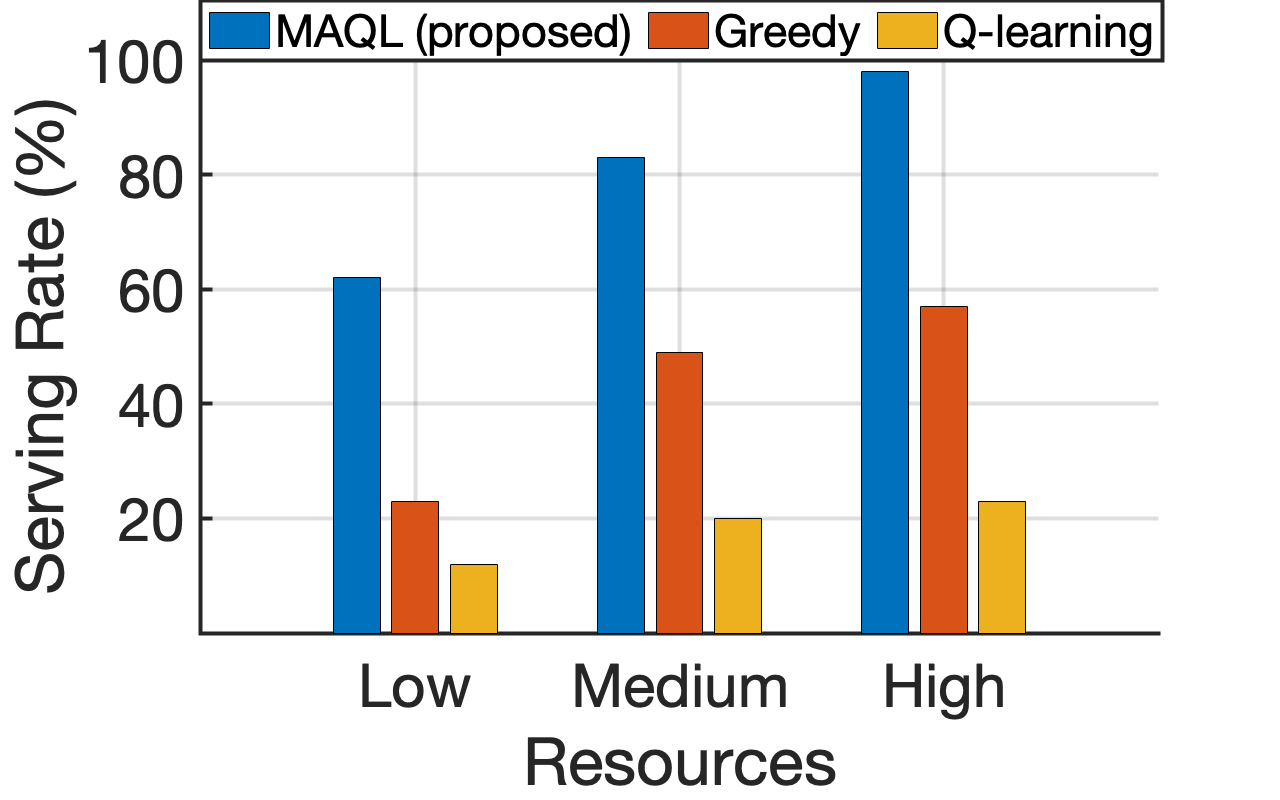}
		%\caption{Single-request context.}
        \caption{Comparison against existing methods in serving rate.}
        \label{fig:sr_benchmarking}
	\end{subfigure}
    ~~% <-- Don't remove these comment to keep horizontal figures
    \begin{subfigure}[b]{0.325\textwidth}
        \centering
	    \includegraphics[width=1\textwidth]{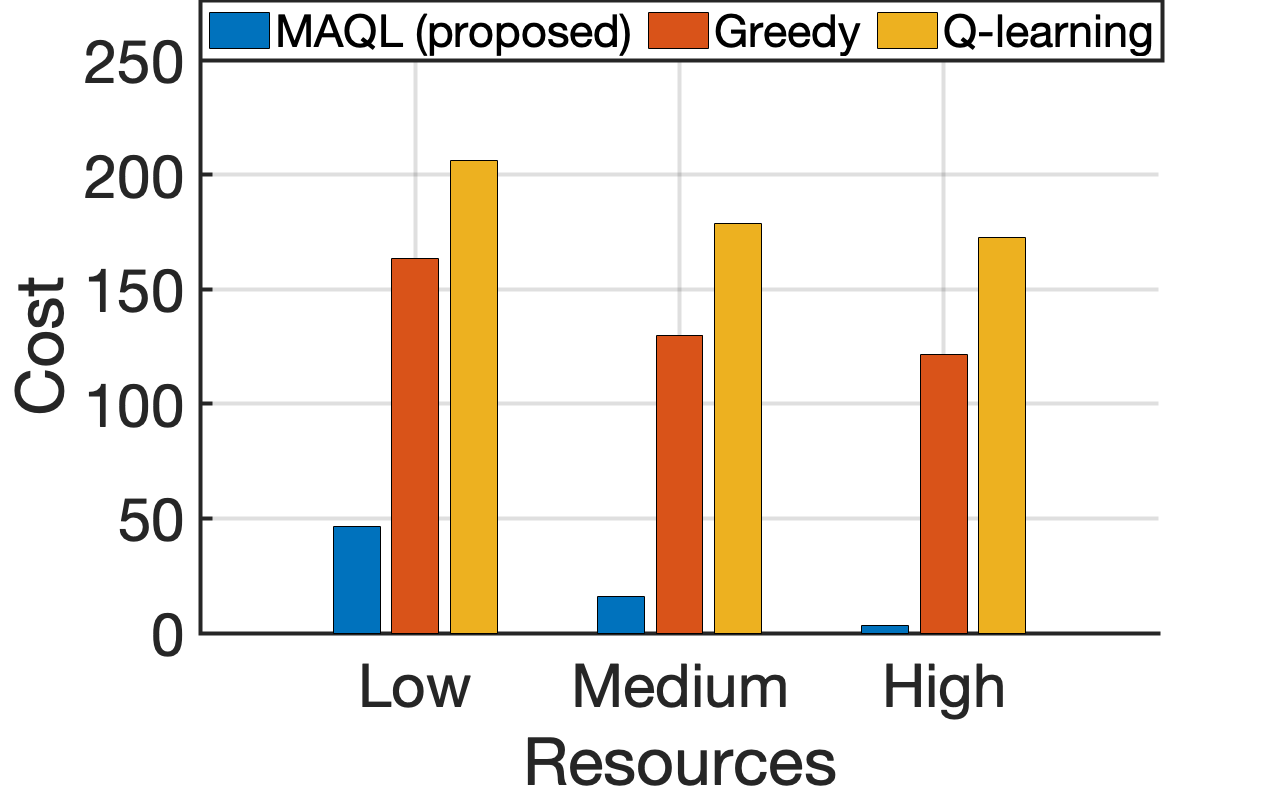}
        %\caption{General context.}
        \caption{Comparison against existing methods in system cost.}
        \label{fig:cost_benchmarking}
	\end{subfigure}
    ~~%
    \captionsetup[subfigure]{
  %labelformat=simple,
  %labelsep=colon,
  textfont=normal, % Caption text will not be bold
  %justification=raggedright,
  labelfont={color=black}, % Label will be blue and bold
  skip=10pt}
    \begin{subfigure}[b]{0.30\textwidth}
        \centering
        %\captionsetup{justification=centering}
		\includegraphics[width=1\textwidth]{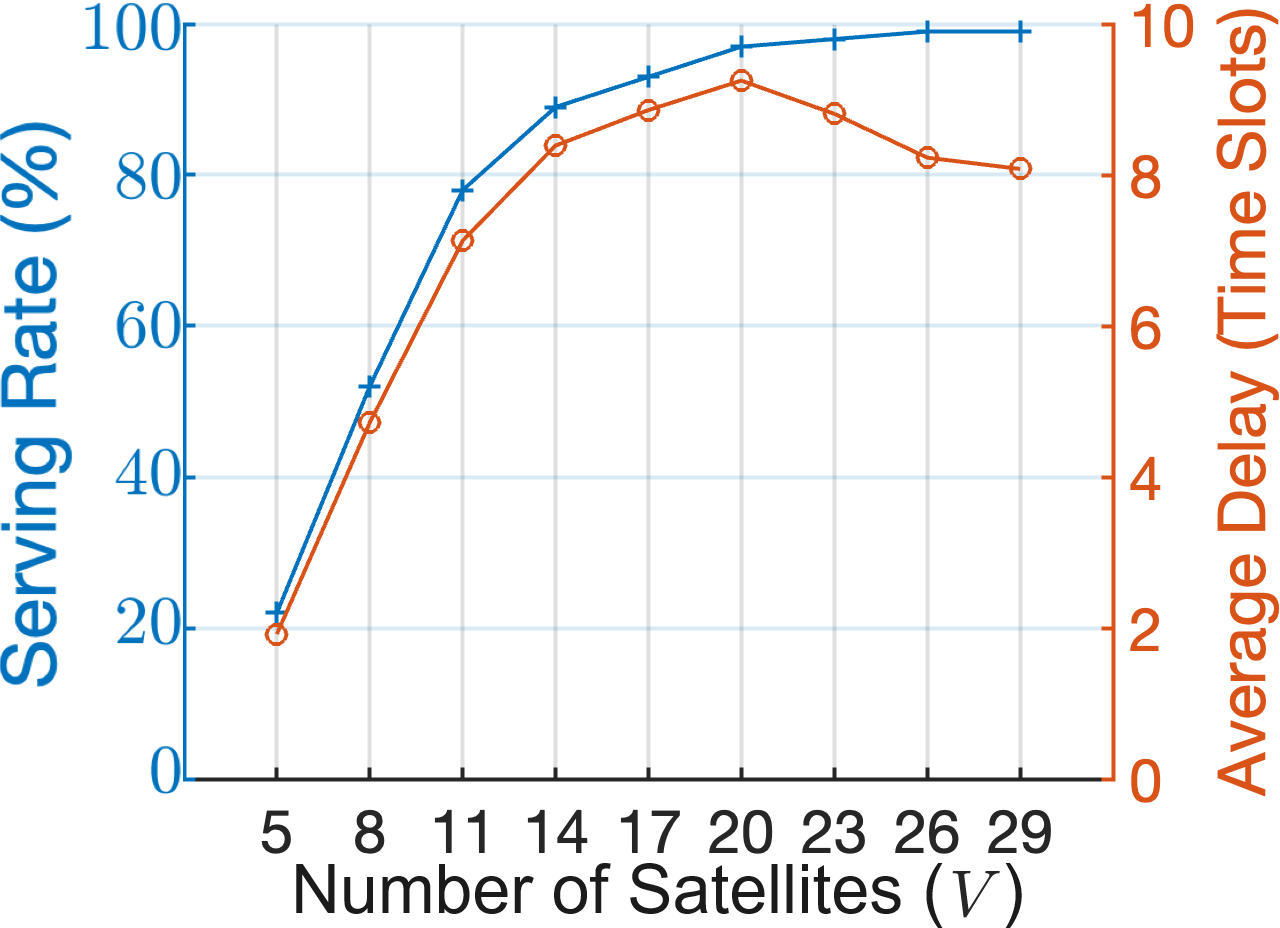}
        %\caption{Impact of $\delta$.}
        \caption{Impact from the number of satellites on the system.}
		\label{fig:Beyond10Sat}
    \end{subfigure}
    \caption{Assessment of the proposed MAQL-based scheme against baseline methods and illustration of the impact from the satellite number on the achieved serving rate and average delay of SFC deployment.}
    \label{fig:maql_vs_baselines}
\end{figure*}

First, we evaluate the efficiency of the MAQL-based VNF placement mechanism. The optimal placement is given by the DP Eq. (\ref{eq:DP}). To calculate the average cost, we compute the discounted long-term cost using Eq. (\ref{eq:longterm_cost_maql}). Additionally, the serving rate is determined by dividing the number of successfully served requests by the total number of requests. The learning rate is $\lambda = 0.1$ and decreases to 0.01 if there is no improvement in a certain period of time. 

\textit{1) Performance against the Optimal solution:} We compare the proposed scheme against the optimal in Fig.~\ref{fig:setup1}. Our simulation setup involves $V=3$ satellites, and a set of available services $\mathbb{H} =\{1,3\}$. The sets of VNFs cached on the satellites are given as $\mathbb{F}_1 = \{1, 2 \}$, $\mathbb{F}_2 = \{2 \}$ and $\mathbb{F}_3 = \{3 \}$, respectively. The active ISLs periods between satellites are $T_{1,2} = 2$, $T_{2,3} = 4$ and $T_{1,3} = 4$ time slots. The number of computational and storage resource units are equal, $R_v = Z_v$, $\forall v \in \mathbb{V}$.  To set the reference point, Fig.~\ref{fig:dp_setup1} presents an analysis on the optimal discounted long-term cost and serving rate achieved through the proposed DP Eq. (\ref{eq:DP}); as expected, the cost decreases and the serving rate increases as the satellites become more resourceful. Next, we define three levels of available satellite resources, \textit{Low}, \textit{Medium} and \textit{High}, where $R_v = Z_v = 4$, 7 and $14$, respectively. Having established the optimal behavior, in Fig.~\ref{fig:sr_setup1} we compare the proposed MAQL approach against it; we observe that the proposed VNF placement solution quickly converges to near-optimal serving rates for all cases. Nonetheless, the more the initial satellite resources, the better the performance as the solution search space becomes broader. Similar observations can be made from Fig. \ref{fig:cost_setup1}, where the long-term system cost achieved after convergence is shown to be close to the optimal. We note here that throughout the experimentation under the Low satellite resources, $59\%$ of the times the placement policy had VNF $2$ executed on satellite $2$ and $41\%$ of the times on satellite $1$; this preference on satellite $2$ is a result of the resource contention for the limited resources of satellite $1$. We should also note that after the one-time offline training of the algorithm, inferring the best policy is instantaneous; in this simple setup, the online execution time is on average $2.4\times 10^{-3}sec$ for MAQL, while the optimal DP-based solution requires $128.57sec$ to calculate a solution, making it unfit for real-time decision making.

%\begin{figure}[h]
%    \centering
%    \begin{subfigure}[t]{0.32\textwidth}
%        \includegraphics[width=1\textwidth]{Figures/Fig5a.png}
%	    \caption{\textcolor{blue}{Serving Rate benchmarking.}}
%	    \label{fig:sr_benchmarking}
%    \end{subfigure}
%    \begin{subfigure}[t]{0.32\textwidth}
%        \includegraphics[width=1\textwidth]{Figures/Fig5b.png}
%	    \caption{\textcolor{blue}{Cost benchmarking.}}
%	    \label{fig:cost_benchmarking}
%    \end{subfigure}
%    \caption{MAQL benchmarking against baselines.}
%    \label{fig:caching_benchmarking}
%\end{figure}

\textit{2) Performance against baselines and impact of the satellite number:} We then assess the proposed algorithms against two baselines and illustrate the impact of a growing number of satellites in Fig. \ref{fig:maql_vs_baselines}. Here, we assume a system with the SFC set $\mathbb{H} = \{4,...,13\}$ and $V=6$ satellites grouped in two subgroups: $\mathbb{V}_1 = \{1, 3, 5\} \subseteq \mathbb{V}$ and $\mathbb{V}_2 = \{2, 4, 6\}\subseteq \mathbb{V}, \mathbb{V}_1 \cup \mathbb{V}_2 = \mathbb{V}$. The active ISL periods are $T_{v,u} = 2$ if $v,u \in \mathbb{V}_1$ or $v,u \in \mathbb{V}_2$, $T_{v,u} = 4$ otherwise. The sets of cached VNFs are given as $ \mathbb{F}_1 = \{1, 2 \}$, $\mathbb{F}_2 = \{3, 4 \}$, $\mathbb{F}_3 = \{5, 6 \}$, $\mathbb{F}_4 = \{7, 8 \}$, $\mathbb{F}_5 = \{9, 10 \}$ and $\mathbb{F}_6 = \{1, 10 \}$. 
%The computational and storage capacities of each satellite are $R_v = Z_v = 2, \forall v \in \mathbb{V}$. 
A brief description of the compared baselines follows:
% \begin{enumerate}
% \item[i)] 

\textit{a)} \textit{Q-learning \cite{9236981}:} we modified this learning VNF placement algorithm for an LSN environment as follows:
    \begin{itemize}
        \item \textit{State}: the requested SFC and VNF for execution.
        \item \textit{Action}: the satellite that will execute the above VNF.
        \item \textit{Reward}: $25(\iota_C + \iota_R)\cdot\exp(2 d / \psi)$,
        where $\iota_C$ and $\iota_R$ are the ratios between the required/available computational and storage resources on the designated satellite, respectively; $d$ is the end-to-end delay of the SFC placement, and $\psi$ is the length of the SFC. Other parameters values are as in \cite{9236981}. 
    \end{itemize}
    We note that this Q-learning-based method is a centralized method that makes a placement decision for an entire SFC in a single time slot, based on the set of active ISLs.
    
\textit{b)} \textit{Greedy \cite{7332796}:} We modify this work's mechanism to fit our LSN environment by matching our requester satellite concept with both the \textit{ingress} and the \textit{egress} switch since the data must be returned to the requester. Based on this, a satellite $v\in \mathbb{V}$ selects another satellite to transfer data by assigning them a cost score. The original cost function is altered as follows: $\mathcal{G}(v, u) = \epsilon_{v,u} + \epsilon_{u,v}$ where $\epsilon_{v,u}$ is the delay for transferring data from satellite $v$ to $u$, which consists of the waiting time for the $(v, u)$-ISL and the transferring time ($1$ time slot). A satellite $v$ will transfer the data to satellite $u$ with the earliest ISL availability (if more than one, it selects randomly uniformly). The VNF placement is performed step by step in each time slot.
% \end{enumerate}
%\begin{figure}
%	\centering
%	%\captionsetup{justification=centering}
%	\includegraphics[width=0.67\columnwidth]{Figures/delay_di2.png}%example_satmobi.png
%	\caption{\textcolor{blue}{Serving Rate and Average Delay per request as the number of available satellites increases.}}
%    \label{fig:Beyond10Sat}
%\end{figure}

To make a fair comparison, none of the three algorithms have knowledge of the pre-installed VNFs at the beginning of the experiment. Figs.~\ref{fig:sr_benchmarking} and \ref{fig:cost_benchmarking} illustrate the results of this benchmarking, in terms of achieved serving rate and cost. The MAQL-based solution outperforms the alternatives by allowing satellites to share learning parameters, enabling cooperative policy updates toward optimal solutions. The Greedy method prioritizes shorter paths while it transfers data between satellites without any VNF caching knowledge, resulting in longer end-to-end delays. Although the high end-to-end delay tolerance in our setting allows data to eventually reach a satellite with the requested VNF cached, stricter tolerances would further degrade the Greedy approach's serving rate. On the other hand, although the plain Q-learning method can eventually learn the VNF caching information, it yields a lower performance compared to the Greedy scheme; this mechanism does not consider the ISL periodicity, meaning that data can be transferred during an inactive ISL, deeming the placements invalid.

Subsequently, we examine the influence of the number of available satellites $V$ on the proposed VNF placement scheme's performance. The satellites are divided into three subgroups: $\mathbb{V}_1$ contains the first $\lceil V/3 \rceil$ satellites, $\mathbb{V}_2$ contains the next $\lceil V/3 \rceil$ satellites, $\mathbb{V}_3$ contains the rest, with $\mathbb{V}_1 \cup \mathbb{V}_2 \cup \mathbb{V}_3 = \mathbb{V}$. ISLs are always active within a subgroup, and intermittent between subgroups with the following periods: $T_{v,u}=2$ if none of $v$ and $u$ belongs to $\mathbb{V}_3$; $T_{v,u}=4$ if either $v$ or $u$ belong to $\mathbb{V}_3$. Each satellite has $R_v=Z_v=3$ units of available resources and is installed with one VNF. We investigate such a limited-resource scenario in order to emphasize the impact of the topology expansion as the number of satellites, $V$, increases. $H=11$ SFCs are considered with $\mathbb{H}=\{4, \ldots, 14 \}$ (details available in Table \ref{tab:SFC_set1}) which comprise $10$ VNFs. The VNF execution time is fixed at 2 time slots, i.e., $d^h_{f,v}=2, \forall h,f,v$. VNFs are selected and installed in the satellites in a round-robin fashion. The results are presented in Fig.~\ref{fig:Beyond10Sat}. %, and hereafter, we will provide more intuition about the behaviors of the two curves in the figure.
%}
There, we observe that in terms of request serving rate, the performance increases significantly (from $22\%$ to almost 100\%) when more satellites are available, due to the increased resource and caching capacity. The right y-axis of Fig. \ref{fig:Beyond10Sat} presents the average end-to-end service delay calculated by dividing the total induced delay by the total number of requests, including successfully served and rejected ones. When $V$ is small compared to the number of VNFs (e.g., $V=3$), there might not be enough VNFs installed in the network to form the entire requested SFC. However, a satellite might still execute an (installed) VNF and transfer the results to another satellite, causing a small delay before the request is rejected (e.g., around 2 time slots for $V=5$). This occurs because the satellites are not aware of the installed VNFs and resources of each other. Naturally, the larger the $V$ is, the more complete sequences of SFCs can be deployed, resulting in an increase on average delay alongside the serving rate. On the other hand, as $V$ increases, $|\mathbb{V}_1|$, $|\mathbb{V}_2|$ and $|\mathbb{V}_3|$ also increase, allowing for more VNFs to be installed in each subgroup. This, in turn, increases the probability for an SFC to be deployed as a whole within a single subgroup where ISLs are always active. This further reduces the deployment time by avoiding the waiting time for ISLs. As a result, after the sweet spot of $V=20$ where the serving rate is maximized, the average delay starts decreasing from $9.25$ to $8.08$ time slots per request.

\textit{3) VNF placement with large-scale service request}: We now extend our analysis to a large-scale service request context, introducing increased randomization in the service request events. In particular, the set of VNFs is $\mathbb{F}=\{1,\ldots,10\}$, and a random requested SFC $h$ is generated via Algorithm \ref{algo:request_generate} where the input distributions $\mathcal{D}_{\text{S}}$ and $\mathcal{D}_{\text{F}}$ are Uniform Distributions. The total number of SFCs in the sample space of Algorithm \ref{algo:request_generate} is computed by: $H=\sum_{\nu=1}^{10} \nu!\binom{10}{l}=9864100$ SFCs. We consider a setup with $V=10$ satellites separated into three subgroups $\mathbb{V}_1=\{1,\ldots,4\}$, $\mathbb{V}_2=\{5,\ldots,8\}$, and $\mathbb{V}_3=\{9,10\}$. The available computational and storage resources are given by $R_v=Z_v=14$ units. We let $L=\underset{h \in \mathbb{H}}{\max}~l_h$ be the maximum length of a requested SFC, and consider three different scenarios where $L=5$, 8, and 10, respectively. Intuitively, the greater the length of a requested SFC, the more resource units and time slots are required for the deployment. Also, we set $D_h=80$ and $d_{f,v}^h=1$ time slots, $\forall h$, as the delay tolerance of SFC and VNF execution duration, respectively. ISLs in the same subgroups are always available. $T_{v,u}=2$ if neither $v$ nor $u$ is in $\mathbb{V}_3$ and $T_{v,u}=4$ if either $v$ or $u$ is in $\mathbb{V}_3$.

\captionsetup[algorithm]{
  labelformat=simple,
  labelsep=colon,
  textfont=normal, % Caption text will not be bold
  justification=raggedright,
  labelfont={color=black, bf}, % Label will be blue and bold
  skip=10pt % Adjust vertical space if needed
}
\begin{algorithm} 
	\caption{Random SFC Request Generation} \label{algo:request_generate}

	\begin{algorithmic}[1]
		%\Procedure{MyProcedure}{}
        \State \textbf{Input:} VNF set $\mathbb{F}$, distributions $\mathcal{D}_{\text{S}}$ and $\mathcal{D}_{\text{F}}$.
        \State \textbf{Output:} $\mathbb{S}$ - the ordered sequence of VNFs.
        %\State \textbf{\textit{Generate SFC's length:}}
        \State Sample a random number $\nu \in \{1,\ldots,|\mathbb{F}| \}$ following distribution $\mathcal{D}_{\text{S}}$.
        %\State \textbf{\textit{Generate VNFs for the SFC:}}
        \State Initialize $\tilde{\mathbb{F}} \leftarrow \mathbb{F}$ and $\mathbb{S} \leftarrow \emptyset$.
        \For{the $f^{\text{th}}$ VNF where $f=1,2,\ldots,\nu$}
        \State Sample a VNF $\mathcal{F} \in \tilde{\mathbb{F}}$ following distribution $\mathcal{D}_{\text{F}}$.
        \State $\mathbb{S} \leftarrow \mathbb{S} \cup \{ \mathcal{F}\}$ and $\tilde{\mathbb{F}} \leftarrow \tilde{\mathbb{F}} \backslash \{ \mathcal{F} \}$.
        \EndFor
        \Return $\mathbb{S}$
		%\EndProcedure
	\end{algorithmic}
 
\end{algorithm}
\begin{figure}
	\centering
	%\captionsetup{justification=centering}
	\includegraphics[width=0.47\columnwidth]{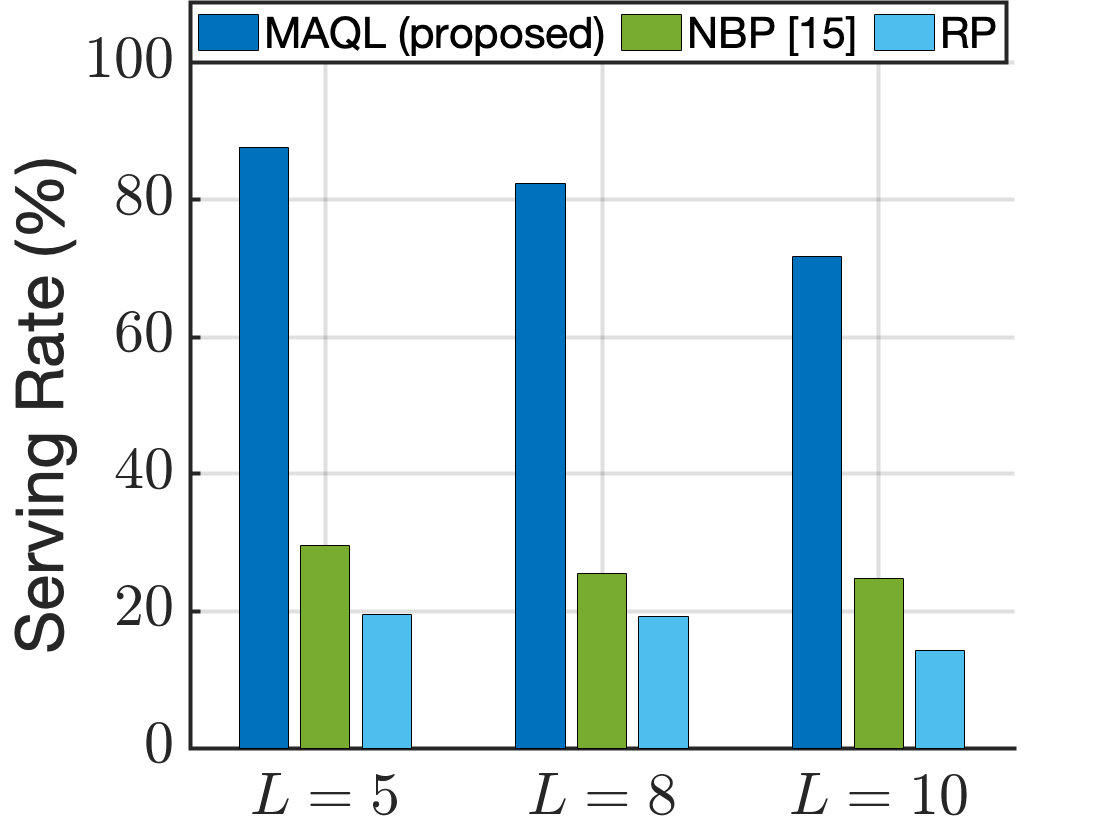}
	\caption{Comparison of the proposed MAQL-based scheme with the NBP \cite{GLTZM_2022} and the RP schemes for different maximum SFC's lengths, $L$.}
    \label{fig:largeSFC_placement}
\end{figure}

We compare the proposed MAQL-based scheme against the Neighbor-Based Placement (NBP) scheme proposed in \cite{GLTZM_2022}. Originally, in \cite{GLTZM_2022}, placement decisions for all VNFs in the requested SFC are made simultaneously, and only available ISLs and resources at the placement time are considered. A request is rejected if no satellite in the sub-network can execute it. To make the comparison fair, we modified the NBP method as follows: satellites can only execute VNFs cached on them, and requests are rejected with a penalty incurred if either the resources are insufficient or the required VNFs are not cached. Additionally, the Random Placement (RP) scheme is also considered as a baseline in which a satellite selects an action uniformly randomly for each request in its buffer.
%We consider the Neighbor-Based Placement (NBP) scheme proposed in \cite{GLTZM_2022} as a benchmark for comparison with the proposed MAQL-based scheme. Originally, in \cite{GLTZM_2022}, the placement decisions for all the VNFs in a requested SFC are made simultaneously. In addition, the NBP scheme works under the constraint that only available ISLs and resources at the placement time are considered in the decision; a satellite can execute a VNF only if its available resources are sufficient. A request is rejected if no satellite in the sub-network can execute it. To make the comparison fair, we modified the NBP method as follows: satellites can only execute VNFs cached on them, and requests are rejected with a penalty incurred if either the resources are insufficient or the required VNFs are not cached. Furthermore, the Random Placement (RP) scheme is also considered as a baseline in which a satellite selects an action uniformly randomly for each request in its buffer.

Our results are in Fig. \ref{fig:largeSFC_placement}. The MAQL-based scheme shows a superior serving rate compared to the NBP scheme. This advantage arises because the proposed approach allows satellites to handle one SFC deployment step per time slot, leading to more effective use of future ISLs and resources. Although the NBP scheme optimally places VNFs under the given constraints, its performance is significantly limited by the requirement that satellites can only execute installed VNFs. This restriction, which relies solely on currently available satellites, severely reduces the number of executable VNFs. In contrast, the MAQL-based framework demonstrates a clear advantage in this context. Besides, the RP scheme, with a lack of strategic planning, shows lower performance compared to the others.

\subsection{BO-based VNF Caching}
In what follows, we evaluate the efficiency of the BO-based VNF caching mechanism. To simplify the evaluation of this component and to avoid any bias potentially induced by the MAQL-based VNF placement, we test the caching mechanism through the following straightforward, greedy VNF placement scheme: if a satellite $v$ has the required VNF cached, it executes it, otherwise, $v$ forwards the request to satellite $u$ with the earliest ISL availability, and which has the required VNF cached. It is assumed that every satellite has complete knowledge of the cached VNFs on every other satellite. For this experimentation family, we first utilize an infrastructure consisting of $V=20$ satellites and $H=2$ SFCs, specifically $h \in \mathbb{H} =\{1,2\}$. We group the satellites into three distinct subgroups as follows: $\mathbb{V}_1 = \{1,...,7\}$,  $\mathbb{V}_2 = \{8,...,14\}$, and  $\mathbb{V}_3 = \{15,...,20\}$, with $\mathbb{V}_1 \cup \mathbb{V}_2 \cup \mathbb{V}_3 = \mathbb{V}$. The ISLs between satellites of the same group are always active. The rest of the active ISL periods are given as: $T_{v,u}=2$ between satellites of $\mathbb{V}_1$ and $\mathbb{V}_2$, $T_{v,u}=5$ between satellites of $\mathbb{V}_1$ and $\mathbb{V}_3$ and $T_{v,u}=3$ between satellites of $\mathbb{V}_2$ and $\mathbb{V}_3$. Each satellite can cache at most two VNFs, i.e., $|\mathbb{F}_v|=2$, $\forall v \in \mathbb{V}$ and their initial resource capacities are equal to $R_v = Z_v = 9$, $\forall v \in \mathbb{V}$.

\begin{figure*}[b]
    \centering
    %\captionsetup[subfigure]{justification=centering}
    %\captionsetup{justification=centering}
	\begin{subfigure}[b]{0.327\textwidth}
        %\captionsetup{justification=centering}
		\includegraphics[width=1\textwidth]{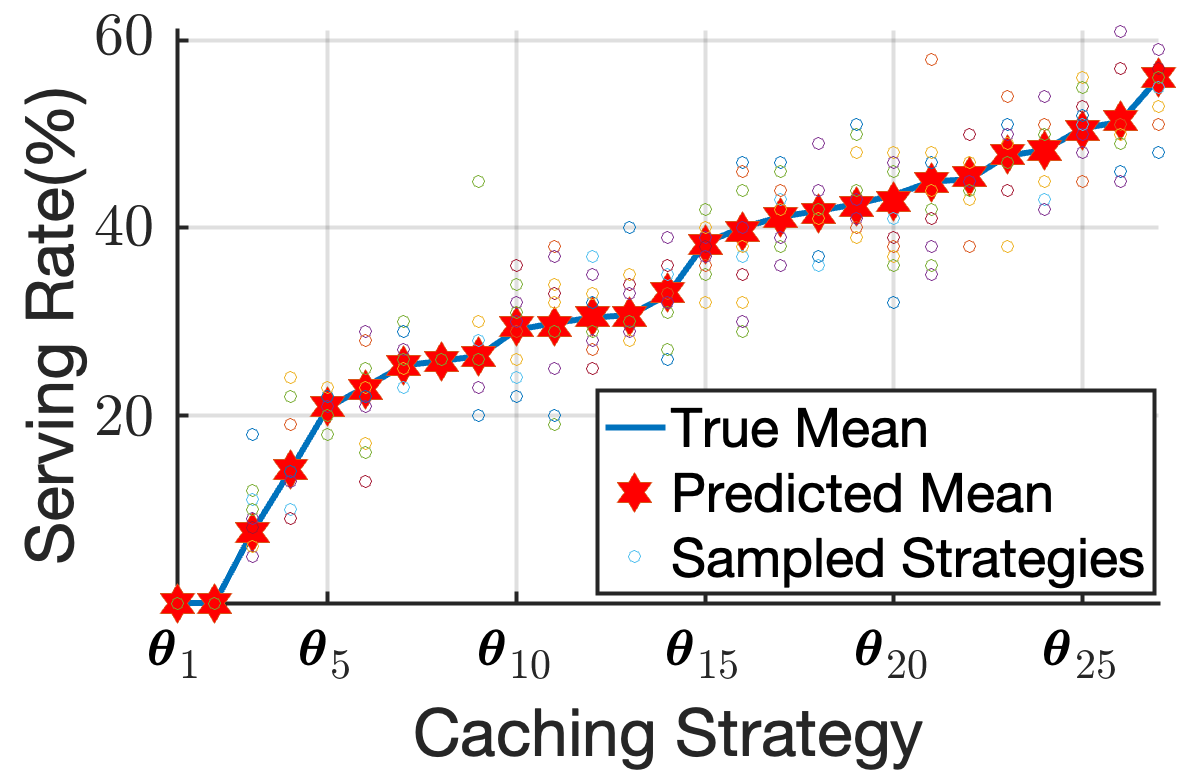}
		\caption{Required number of evaluations without $I(\cdot)$.}
		\label{fig:non_acquif}
	\end{subfigure}
    % <-- Don't remove these comment to keep horizontal figures
    \begin{subfigure}[b]{0.327\textwidth}
	    \includegraphics[width=1\textwidth]{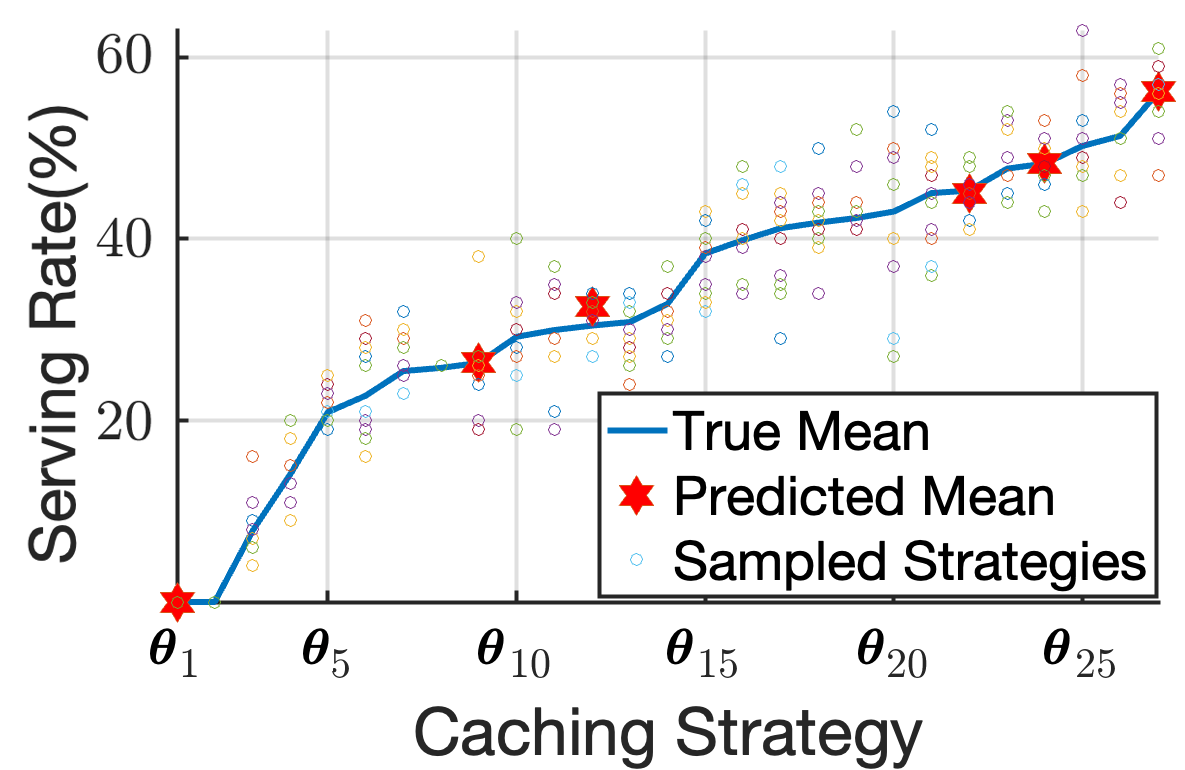}
		\caption{Required number of evaluations with $I(\cdot)$.}
		\label{fig:acquif}
	\end{subfigure}
    \begin{subfigure}[b]{0.29\textwidth}
	    \includegraphics[width=1\textwidth]{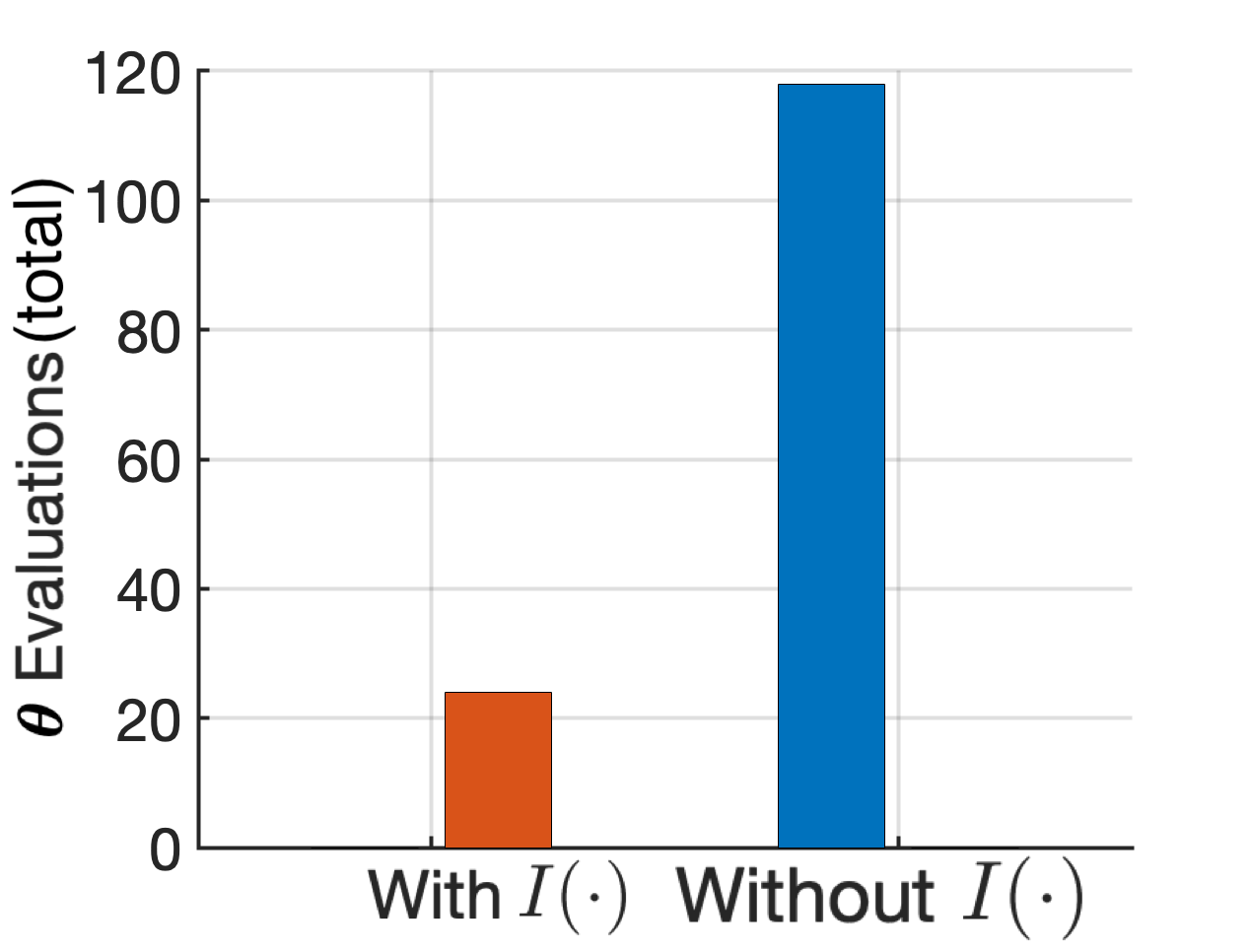}
		\caption{Effectiveness of the employment of $I(\cdot)$.}
		\label{fig:bench-acquif}
	\end{subfigure}
    \caption{Demonstration of acquisition function $I(\cdot)$'s impact on determining potentially optimal caching strategies.}
    \label{fig:acquif_role}
\end{figure*}

\textit{1) Performance impact of Acquisition Functions:} We study the impact of incorporating an acquisition function $I(\cdot)$ in Fig. \ref{fig:acquif_role}, by closely examining three satellites, one from each group $\mathbb{V}_1$, $\mathbb{V}_2$ and $\mathbb{V}_3$. We employ the PI acquisition function, Eq. \eqref{eq:PoI_func}, in our proposed BO-based caching method to perform caching on three VNFs, $\mathbb{S}_1 \cup \mathbb{S}_2 = \{1,2,3\}$. Given that $|\mathbb{F}_v|=2$, $\forall v \in \mathbb{V}$, this allows for choosing between three unique caching sets $\mathbb{F}_v$ for each satellite, which gives us a search space of $\boldsymbol{\Theta} = 3^3 =27$ caching strategies. In the case where no $I(\cdot)$ is used, $\boldsymbol\theta_{\sf max}$ is sampled randomly from $\boldsymbol{\Theta}$. The results of this case are illustrated in Fig. \ref{fig:non_acquif}, where we observe that the GP model's predictions align closely with the actual average serving rates represented on the x-axis for each caching decision. This emphasizes that the GP serves as a suitable surrogate model for our problem. However, without the guidance of an acquisition function, all the potential cache placements are subjected to evaluation, resulting in a considerable increase in execution time. 
Contrarily, when $\boldsymbol\theta_{\sf max}$ is calculated through an acquisition function as in Eq. (\ref{eq:nextpoint_acquif}), the number of evaluations required is significantly reduced, and the outcome approaches the optimal solution, showcasing a more efficient operation, as shown in Fig. \ref{fig:acquif}. For visualization purposes, the caching strategies in Figs. \ref{fig:non_acquif} and \ref{fig:acquif} are sorted based on the true mean serving rate achieved with them.  In Fig. \ref{fig:bench-acquif}, we illustrate the benefits of using an acquisition function; this reduction of $80\%$ in the number of evaluations needed directly translates to immense savings in real-world execution time. Specifically, in our setup, evaluating every potential caching strategy translated to $73.16sec$ total execution time, whereas the same optimal caching strategy was pinpointed through the acquisition function in just $14.88sec$.

\begin{figure*}[b]
    %\centering
    %\captionsetup[subfigure]{justification=centering}
    %\captionsetup{justification=centering}
	\begin{subfigure}[b]{0.32\textwidth}
        \centering
        %\captionsetup{justification=centering}
		\includegraphics[width=1\textwidth]{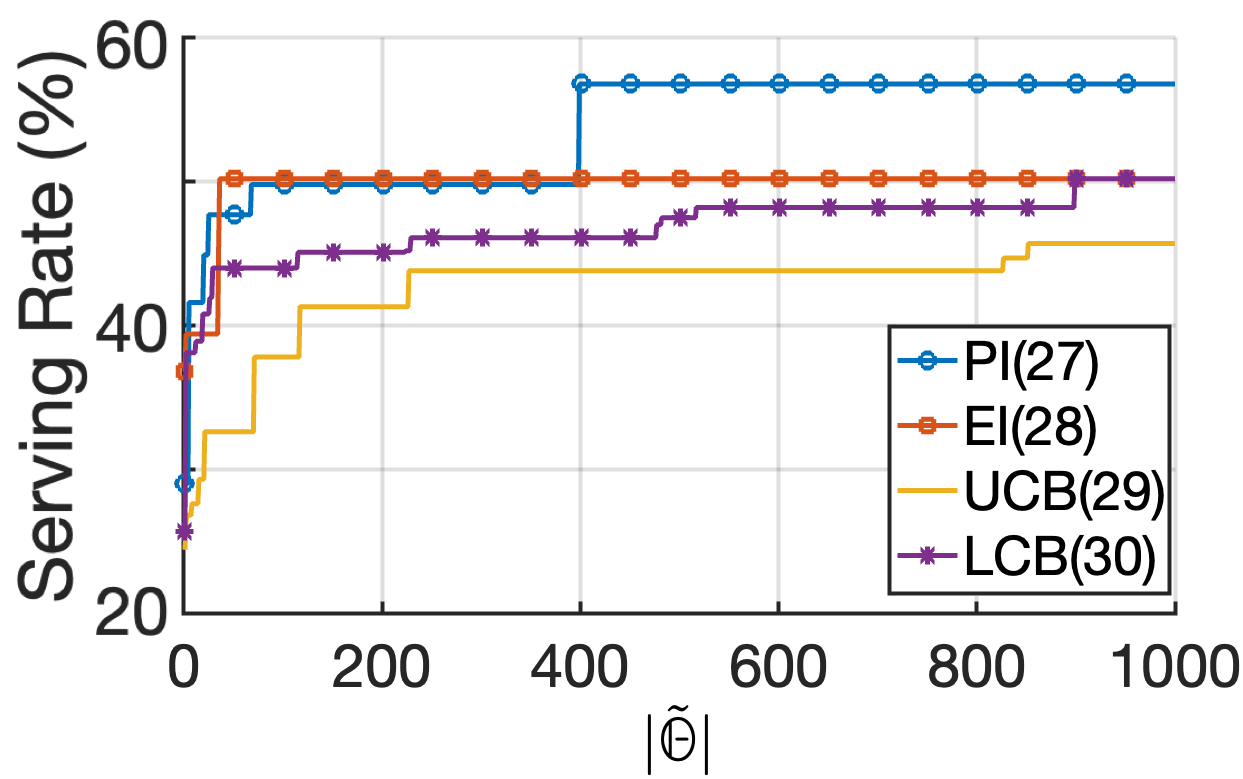}
		%\caption{Single-request context.}
        \caption{Achieved serving rate vs training set size.}
        \label{fig:bestoverites}
	\end{subfigure}
    % <-- Don't remove these comment to keep horizontal figures
    \begin{subfigure}[b]{0.35\textwidth}
        \centering
        \includegraphics[width=1\textwidth]{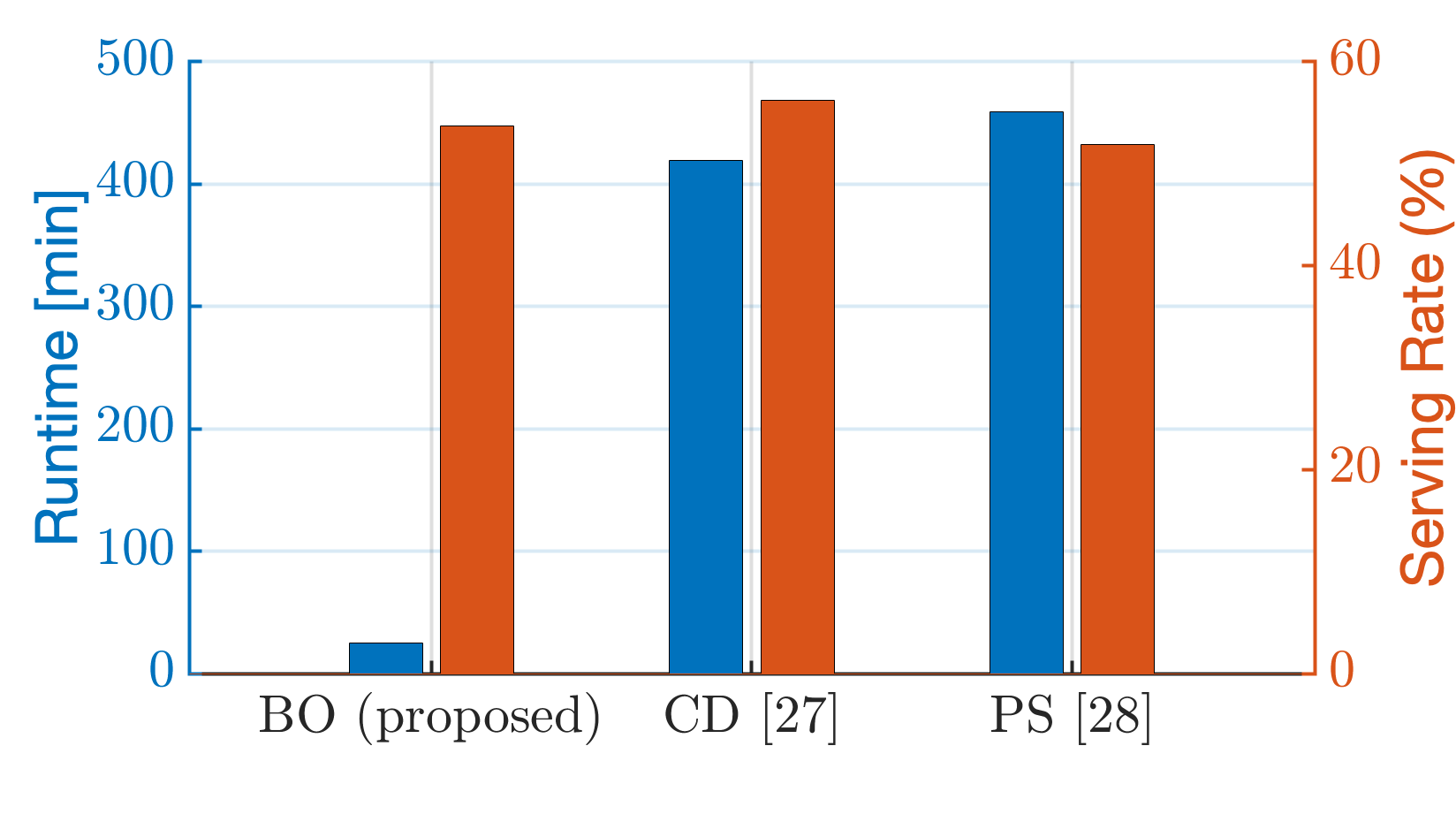}
        %\caption{General context.}
        \caption{Comparison against baseline methods.}
        \label{fig:4methods}
	\end{subfigure}
    \captionsetup[subfigure]{
  %labelformat=simple,
  %labelsep=colon,
  textfont=normal, % Caption text will not be bold
  %justification=raggedright,
  labelfont={color=black}, % Label will be blue and bold
  skip=10pt}
    \begin{subfigure}[b]{0.315\textwidth}
        \centering
        %\captionsetup{justification=centering}
		\includegraphics[width=1\textwidth]{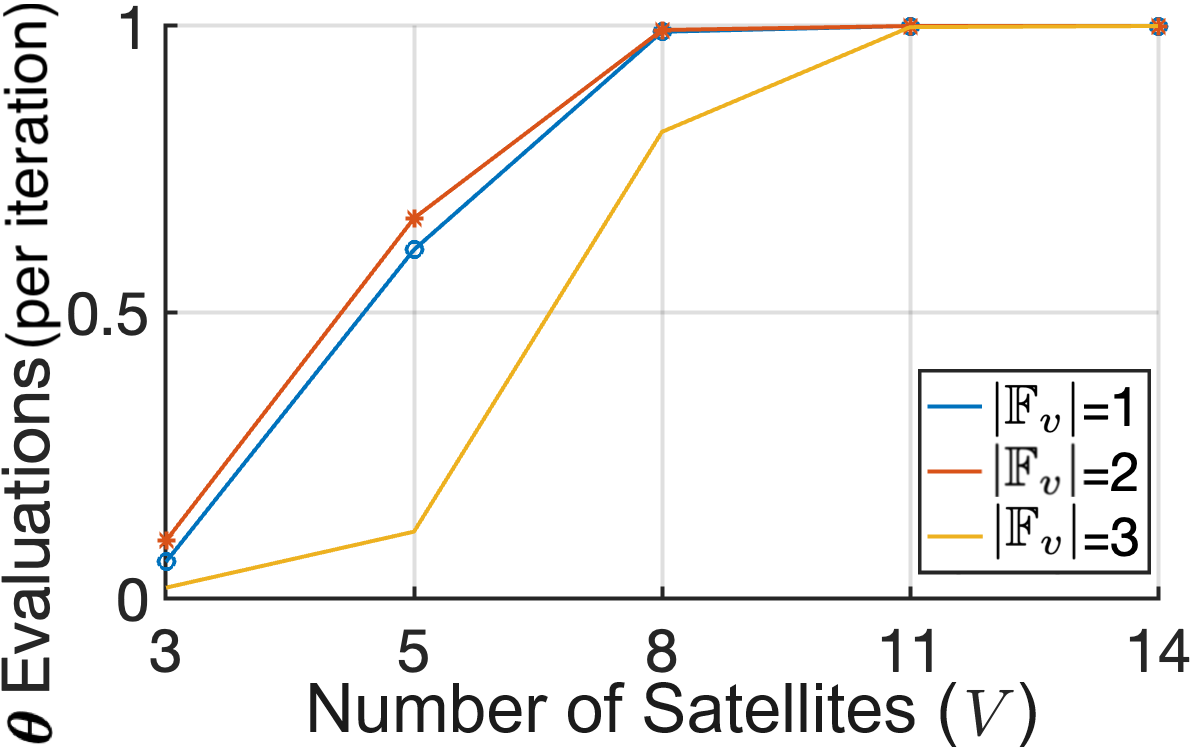}
        %\caption{Impact of $\delta$.}
        \captionsetup{width=1.03\textwidth}
        \caption{Number of $\boldsymbol\theta$ evaluations per iteration.}
		\label{fig:runningtime_satnum}
    \end{subfigure}
    \caption{Evaluations of the proposed BO-based scheme in terms of achieved serving rate and execution delay. }
    \label{fig:vnf-bench}
\end{figure*}

\textit{2) Performance impact of Acquisition Function type, cache size and comparison against baselines:} In Fig. \ref{fig:vnf-bench}, we compare the performance of the four acquisition functions by analyzing the serving rate achieved as the training set size $|\tilde{\boldsymbol{\Theta}}|$ increases with each iteration of Algorithm \ref{algo:BO}.
%For the next experiment, we examine the performance of the method when the four different types of acquisition functions introduced in Section \ref{sec:BOA} are used, i.e., PI, EI, UCB, and LCB. 
%The y-axis of Fig.~\ref{fig:bestoverites} indicates the highest serving rate achieved over an incremental number of iterations. 
The superior performance of the PI function, Eq. \eqref{eq:PoI_func}, as illustrated in Fig. \ref{fig:bestoverites},  suggests that in the VNF caching problem, exploration and prioritizing immediate improvements are dominant factors. PI places a strong emphasis on exploring regions of the search space where improvements are likely, and this approach evidently proved advantageous. On the other hand, the EI function, Eq. \eqref{eq:EI_func}, which also prioritizes exploration but to a slightly lesser extent, provided consistently good results, though subpar compared to PI. Contrarily, the UCB Eq. \eqref{eq:UCB_func} and LCB Eq. \eqref{eq:LCB_func} functions emphasize exploration by considering upper and lower bounds on the evaluated function, but they do not directly measure the likelihood of improvement over the current best value. This undermines the significance of identifying areas with the potential for caching policy improvements, and this is reflected in the mediocre performance. This improvement of the serving rate in this function follows a step function pattern, as the best serving rate achieved is improved only when a better caching solution is found. 
Following, in Fig.~\ref{fig:4methods} we demonstrate the results of benchmarking the proposed VNF caching mechanism (BO) against two  baselines from the literature: 
%\textit{Coordinate Descent (CD)} to \cite{wright_recht_2022}, \textit{Pattern Search (PS)} \cite{BOGANI2009283}, and \textit{Random Search (RS)}. We provide a brief description of these three algorithms:
% \begin{enumerate}
% \item[i)]

\textit{a)} \textit{Coordinate Descent (CD) \cite{wright_recht_2022}}: initialize the caching strategy $\boldsymbol\theta$ and $i=0$. In each iteration do:
\begin{enumerate}[label=(\roman*)]
    \item[(i)] Produce $|\bigcup^H_{h=1} \mathbb{S}_h|$ different caching strategies $\boldsymbol\theta'$ by alternatively assigning the $i^{\text{th}}$ component of $\boldsymbol\theta$ with the VNF indices $1, \ldots, |\bigcup^H_{h=1} \mathbb{S}_h|$. Then, evaluate $M_{\Pi}(\mathbf{\boldsymbol\theta'})$ for every obtained $\boldsymbol\theta'$.
    \item[(ii)] update $\boldsymbol\theta$ to the $\boldsymbol\theta'$ with the maximum evaluation. 
    \item[(iii)] $i \leftarrow i+1$.
    \item[(iv)] Return $\boldsymbol\theta$ if termination criteria met, else \textit{go to} (i).
\end{enumerate}
\textit{b)} \textit{Pattern Search (PS) \cite{BOGANI2009283}}: initialize $\boldsymbol\theta$ randomly. In each iteration do:
\begin{enumerate}[label=(\roman*)]
    \item[(i)] generate neighbor strategies $\boldsymbol\theta' = \left\{\theta^{'}_i | \theta^{'}_i \in \mathbb{F} \right\}$; $\boldsymbol\theta'$ is a neighbor of $\boldsymbol\theta$ if there exists indices $j \in \{1, \ldots, N \}$ and $k \in \{1, \ldots, |\mathbb{F}|\}$ such that $\theta_j = \mathcal{F}_{k}$, $\theta_j^{'} = \mathcal{F}_{k+1}$, and $\theta_i^{'} = \theta_i$ for $i \ne j$. Evaluate $M_{\Pi}(\mathbf{\boldsymbol\theta'}),$ for every obtained $\boldsymbol\theta'$.
    \item[(ii)] update $\boldsymbol\theta$ to the $\boldsymbol\theta'$ with the maximum evaluation. 
    \item[(iii)] Return $\boldsymbol\theta$ if termination criteria met, else \textit{go to} (i).
\end{enumerate}
% \end{enumerate}
% \subsubsection{Random Search} sample randomly uniformly caching strategies $\boldsymbol\theta$. Evaluate them using $M_{\Pi}(\mathbf{\boldsymbol\theta})$. Select the one with the maximum evaluation.
    
\comment{We establish a common termination condition for all algorithms at $100$  consecutive iterations 
%(or equivalent to $10^4$ time slots) 
without an improvement in the serving rate. 
%In Fig.~\ref{fig:4methods}, the left y-axis depicts the highest serving rate achieved upon termination of these algorithms, while the right y-axis displays the total running time in minutes. 
The results depicted in Fig.~\ref{fig:4methods} indicate that the BO, CD, and PS schemes offer comparable serving rates upon completion. However, the BO scheme demonstrates a superior ability to strike a balance between maximizing the objective function and maintaining a low execution time compared to the other candidates. This happens because the other baselines need to invoke the objective function and evaluate all the potential strategies without any insight. In contrast, the proposed scheme can predict whether a new caching strategy would result in an improvement by using the much less computationally expensive acquisition function. Finally, in Fig. \ref{fig:runningtime_satnum} we briefly demonstrate the scalability of the BO-based algorithm. Since the objective function evaluation time accounts for the majority of the running time of the suggested caching algorithm, we opted to show the average $M_\Pi(\boldsymbol\theta)$ evaluations per iteration (lines $6-11$ in Algorithm \ref{algo:BO}) as the size of the infrastructure grows, for different cache sizes $|\mathbb{F}_v|$.}
The results depicted in Fig.~\ref{fig:4methods} indicate that the BO, CD, and PS schemes offer comparable serving rates upon completion. However, the BO scheme demonstrates a superior ability to strike a balance between maximizing the objective function and maintaining a low execution time compared to the other candidates. This is due to the ability of the presented framework in predicting the potential improvement of an unevaluated strategy. Finally, in Fig. \ref{fig:runningtime_satnum} we briefly demonstrate the scalability of the BO-based algorithm. Since the objective function evaluation time accounts for the majority of the running time, we opted to show the average $M_\Pi(\boldsymbol\theta)$ evaluations per iteration (lines $6-11$ in Algorithm \ref{algo:BO}) as the size of the infrastructure grows, for different cache sizes $|\mathbb{F}_v|$.

Next, we consider $\mathbb{H} = \{1,2,4\}$ that results in 4 VNFs to be cached. Additionally, satellite subgroups are formed as follows: $\mathbb{V}_1$ contains the first $\lceil V/3 \rceil$ satellites, $\mathbb{V}_2$ the next $\lceil V/3 \rceil$ satellites and $\mathbb{V}_3$ contains the rest. The ISL periodicity between the logical subgroups remains the same as defined at the beginning of this subsection. The figure suggests that the evaluation rate increases monotonically as the infrastructure size increases. This is consistent with the fact that the total number of caching strategies is given by $\prod_{v=1}^V \binom{|\cup^H_{h=1} \mathbb{S}_h|}{|\mathbb{F}_v|}$. In addition, since the number of possible VNF subsets to cache on a satellite is given by the binomial coefficient $\binom{4}{|\mathbb{F}_v|}$ we see that more evaluations are required when $|\mathbb{F}_v|=2$ and $|\mathbb{F}_v|=1$ and less for $|\mathbb{F}_v| = 3$.
%whose value increases as $\mathbb{F}_v = 3$, 1 and 2, respectively. As a result, the evaluation rate grows at the fastest pace as $\mathbb{F}_v=2$, and the lowest pace as $\mathbb{F}_v = 3$.

\textit{3) Assessing the combined performance of the MAQL-based VNF-Placement \& the BO-based VNF Caching methods:} In this simulation, we compare the proposed MAQL-based scheme with the NBP scheme \cite{GLTZM_2022}. For both schemes, the subsets of installed VNFs on the satellites are selected via the proposed BO-based algorithm. The setup in this experiment is as follows: $\mathbb{H} = \{ 1, 2\}$ and the network consists of $V=3$ satellites with $T_{1,2} = 2$, $T_{1,3}=4$ and $T_{2,3}=4$ being the periods of ISLs between the satellite pairs. The satellite resources and VNF execution time are given as $R_v=Z_v=14,$ and $d^h_{fv}=2, \forall v \in \mathbb{V}$. %The rest of the parameters remain as in Table \ref{tab:defaultval}.}

\begin{figure}[H]
	\centering
	%\captionsetup{justification=centering}
	\includegraphics[width=0.45\columnwidth]{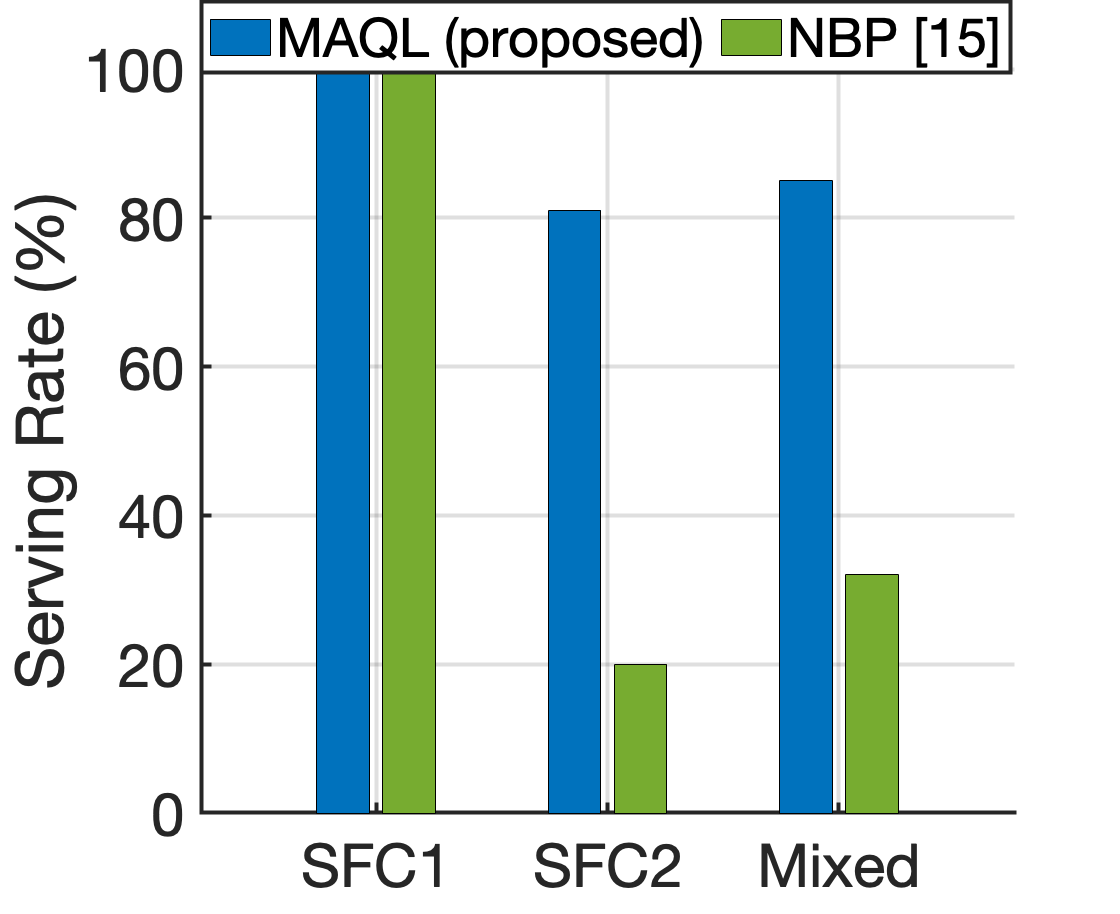}%example_satmobi.png
	\caption{Comparison between the proposed MAQL-based approach and the NBP \cite{GLTZM_2022} schemes, under the BO-based VNF caching method.}
    \label{fig:BOMAQL_Viterbi}
    %\vspace{-10pt}
\end{figure}

The results in terms of achieved serving rate are presented in Fig. \ref{fig:BOMAQL_Viterbi}; ``SFC 1'' stands for the case when SFC 1 is the only available service. Here, the BO-based caching results in $\mathbb{F}_v=\{2, 3\}, \forall v \in \mathbb{V}$ for both the MAQL and NBP methods, which can be straightforwardly confirmed as the optimal caching strategy as SFC 1 comprises only VNF 2 and 3. In this case, every request can be handled by satellites individually, hence, both methods serve $100\%$ of requests. Similarly, ``SFC 2'' stands for the case where SFC 2 is the only available service and ``Mixed'' when both SFCs are available and either is requested with equal probability of $0.5$. In these cases we observe 
%a performance gap between the two compared methods in these cases, with 
that the proposed method prevails due to the limitation of NBP which considers only satellites with currently active ISLs. The superiority of exploiting the periodic movements of satellites to perform the placement using the entire LSN in the case of the proposed MAQL-based solution is evident.

\textit{4) VNF caching with large-scale service request}: In this simulation, we examine the impact of cache size in a large-scale service request setting. We consider the same simulation setup described in Subsection \ref{subsec:simu_maql}-3 where a total of $H=9864100$ SFCs can be requested. We make the following modifications in the setup: $V=12$ satellites that form three subgroups with four satellites in each one and $T_{v,u}=10$ for satellites $v$ and $u$ belonging in different subgroups. To the best of our knowledge, none of the existing works in the literature have proposed a practical distribution for the popularity of VNFs. Therefore, to demonstrate the impact of the VNF popularity imbalance on the caching performance, we assume that the popularity of VNFs, i.e., $\mathcal{D}_{\text{F}}$, follows Zipf distribution \cite{saichev2009theory}. This choice is based on Zipf's skewness factor, $\chi$, which effectively models varying levels of popularity imbalance. Particularly, we consider low and high popularity imbalance levels defined by $\chi=1.5$ and $\chi=4$, respectively. The requested SFCs are generated through Algorithm \ref{algo:request_generate}.

\begin{figure}[H]
	\centering
	%\captionsetup{justification=centering}
	\includegraphics[width=0.5\columnwidth]{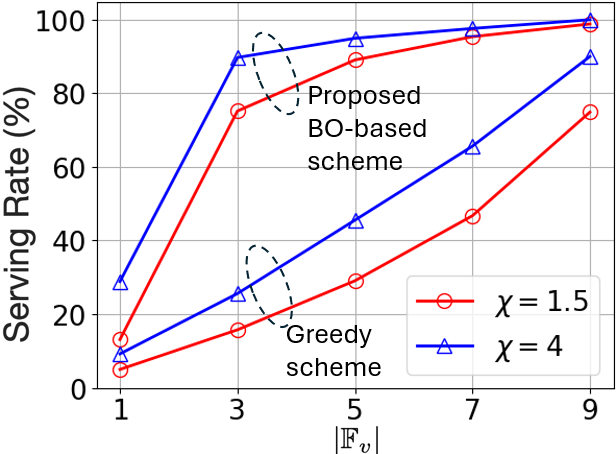}
	\caption{Assessment of the proposed BO-based caching method against the Greedy scheme where VNF popularity following a Zipf distribution with skewness factor $\chi$.}
    \label{fig:largeSFC_cache}
\end{figure}

The results are illustrated in Fig.~\ref{fig:largeSFC_cache} where both methods achieve improved performance as the caching capacity $|\mathbb{F}|_v$, $\forall v\in \mathbb{V}$, increases. We compare the proposed BO-based scheme against a Greedy Caching scheme, where the popularity of VNFs is known and VNFs are cached in the descending order of their popularity. Since satellites operate independently and are not aware of the cached VNF subsets of each other, the greedy scheme, with a lack of diversity in its caching decisions, results in the same cache placement for every satellite. The presented framework can effectively capture the popularity and diversity aspects of VNFs for an effective caching strategy, as demonstrated in Fig.~\ref{fig:largeSFC_cache}. Besides that, the bias of VNF popularity is also shown to have a significant impact with both schemes performing better when the bias defined by $\chi$ increases from $1.5$ to $4$.

\section{Conclusion} \label{sec:conclusion}

In this work, we have tackled the SFC placement problem in LSNs, aiming at optimizing long-term system performance. We have first developed an optimal offline service placement policy using a DP equation, but its high computational complexity, extensive statistical information,  and centralized nature presented challenges for online use. To overcome these, we have proposed a cooperative MAQL-based approach with a parameter-sharing mechanism to manage the non-stationary satellite environment. Additionally, recognizing the dependence of SFC deployment on the pre-installed/cached VNFs on each satellite, we have incorporated a BO-based VNF caching scheme to maximize the request serving rate, where an acquisition function has guided the search towards potentially optimal caching strategies. Simulations have demonstrated that our proposed framework outperforms the most recent approaches introduced in the literature. Future work will focus on scaling the framework to handle multiple requests and exploring a distributed approach to enhance the BO-based method, particularly for applications in LEO satellite mega-constructions.

%%%
\comment{In this work, we have dealt with the SFC placement problem for LSNs in a discrete-time stochastic control framework, with the goal of optimizing long-term system performance. To this end, we have first devised an optimal offline service placement policy by formulating a DP Equation. Due to its high computational complexity and other constraints, solving the DP Equation online could be an unrealistic cumbersome task. To overcome the problem, we have presented a cooperative MAQL-based VNF placement approach. To accommodate the non-stationary nature of the satellite environment, which hampers convergence, we have promoted satellites to share learned parameters. Additionally, as the SFC deployment process depends on the set of VNFs pre-installed/cached on each satellite, we have complemented the mechanism with a BO-based VNF caching scheme, aiming at maximizing the serving request rate. Finally, we have presented experimental results demonstrating the convergence and performance of the MAQL-based method, benchmarking it against the optimal solution and two other approaches. Our BO-based VNF caching scheme has effectively maximized the serving rate, as shown through extensive simulations and benchmarks. Combined, our placement and caching solutions have been shown to outperform a recent work from the literature.}

\comment{For future work, we plan to address the problem in more realistic scenarios where each satellite has a certain failure probability, necessitating a more robust framework. Additionally, the BO-based approach's performance can be enhanced through hyperparameter tuning. A promising approach is a distributed BO framework, where BO methods run on multiple nodes with different hyperparameters. The VNF subsets suggested by these nodes can then be combined using a trainable weighting mechanism.}

%%%
\comment{In this work, we have dealt with the SFC placement problem for LSNs in a discrete-time stochastic control framework, with the goal of optimizing long-term system performance. To this end, we have first devised an optimal offline service placement policy by formulating a DP Equation. Due to its high computational complexity, the requirement of in-advance service request probability knowledge, and the need for a centralized implementation, solving the DP Equation online could be an unrealistic cumbersome task. To overcome the aforementioned challenges, we have presented a cooperative MAQL-based VNF placement approach. To accommodate the non-stationary nature of the satellite environment, which hampers convergence, we have promoted satellites to share learned parameters. Additionally, as the SFC deployment process depends on the set of VNFs pre-installed/cached on each satellite, we have complemented the mechanism with a BO-based VNF caching scheme, aiming at maximizing the serving request rate. Finally, we have presented experimental results that demonstrate the convergence and performance of the MAQL-based method, benchmarking it against the optimal solution and two other approaches from the literature. We have showcased the effectiveness of our proposed BO-based VNF caching scheme in maximizing the serving rate, again through extensive simulation and benchmarking against baselines. The efficacy of the proposed placement and caching solutions combined is additionally shown to outperform a recent work from the literature. In practical scenarios, the proposed MAQL framework can be deployed in a distributed manner where each satellite behaves as an independent controller. In contrast, the BO-based scheme can be deployed centrally with instructions and commands provided by a satellite base station unit on the ground.
%Furthermore, our results offer deeper insights into the role and impact of the acquisition function.

In future work, we plan to investigate the solution to the problem under the more realistic assumption of bulk service request arrival in the system. In addition, we are committed to explore options for optimizing the position of the satellites themselves, by introducing their mobility as an action in the reinforcement learning formulation.}

\bibliographystyle{IEEEtran}
\bibliography{IEEEabrv,DNK}

\comment{
    \begin{IEEEbiography}[{\includegraphics[width=1in,height=1.25in,clip,keepaspectratio]{Revision_Round3/Authors_Photos/khai_photo.png}}]%
{Khai Doan}
    received his PhD in Information Systems Technology and Design from Singapore University of Technology and Design (SUTD), Singapore, in 2020, under SUTD President Graduate Fellowship. Presently, he is working at a Research Professor position in the School of Electrical Engineering, Korea University, Seoul, South Korea. His research interests include edge computing, machine learning, and satellite networks.
\end{IEEEbiography}

	\begin{IEEEbiography}[{\includegraphics[width=1in,height=1.25in,clip,keepaspectratio]{Revision_Round3/Authors_Photos/Marios.jpg}}]%
{Marios Avgeris}
    received the Diploma degree in electrical engineering in 2016 and the Ph.D. degree in electrical engineering from the National Technical University of Athens in 2021. He has been a Postdoctoral Fellow with the Department of Systems and Computer Engineering, Carleton University and the Software and Information Technology Engineering department at the École de Technologie Supérieure (ÉTS) since 2022. His research interests include edge and cloud computing, computational offloading, NFV placement, and network optimization and management.
\end{IEEEbiography}

	\begin{IEEEbiography}[{\includegraphics[width=1in,height=1.25in,clip,keepaspectratio]{Revision_Round3/Authors_Photos/leivadeas.jpg}}]%
{Aris Leivadeas}
    (S'12-M'15-SM'21) is currently an Associate Professor with the Department of Software and Information Technology Engineering at the École de Technologie Supérieure (ETS), Montreal, Canada. From 2015 to 2018 he was a postdoctoral fellow in the Department of Systems and Computer Engineering, at Carleton University, Ottawa Canada. In parallel, Aris worked as an intern at Ericsson and collaborated with Cisco in Ottawa, Canada. He received his diploma in Electrical and Computer Engineering from  the University of Patras in 2008, the M.Sc. degree in Engineering from King’s College London in 2009, and the Ph.D degree in Electrical and Computer Engineering from the National Technical University of Athens in 2015. His research interests include Network Function Virtualization, Network Automation, Cloud and Edge Computing, IoT, and network optimization and management. He received the best paper award in ACM ICPE’18 and ’23, IEEE iThings '21, and the best presentation award in IEEE HPSR’20.
\end{IEEEbiography}

	\begin{IEEEbiography}[{\includegraphics[width=1in,height=1.25in,clip,keepaspectratio]{Revision_Round3/Authors_Photos/IoannisLambadaris.jpg}}]%
{Ioannis Lambadaris}
    was born in Thessaloniki, Greece. He received a diploma in Electrical Engineering from the Polytechnic School of the Aristotle University of Thessaloniki in 1984. He was a recipient at a Fulbright Fellowship (1984-1985) for graduate studies in USA. He received a  M.Sc. degree in Engineering from Brown University, Providence, RI, USA in 1985 and a Ph.D. degree in Electrical Engineering from the University of Maryland, College Park, MD, USA in 1991. 

He was employed as a research associate at Concordia University, Montreal, Quebec, Canada, in 1991-1992. He joined the Department of Systems and Computer Engineering in Carleton University in September 1992. Currently, he is a Chancellor's professor in the same department. While at Carleton he received the Premiere Research Excellence Award (2000), and the Carleton University Research Excellence Award (2000-2001) for his research achievements in the area of modeling and performance analysis of computer networks. In 2020 Prof. Lambadaris was awarded the Ericsson Chair in 5G Wireless Research (https://carleton.ca/ericsson/ericsson-chair-in-5g-wireless-research/)

Professor Lambadaris' interests lie in the area of applied stochastic processes, stochastic control, queuing theory and their application for modeling/simulation and performance analysis of computer communication networks. He has numerous contributions in the areas of quality of service (QoS) control for IP networks, resource allocation in optical networks, and optimal routing and flow control in ad-hoc wireless systems. His recent research focus is in the areas of hardware and software solutions for mobile applications and platforms with a focus on biomedical, remote monitoring, and security applications. 

\end{IEEEbiography}

\begin{IEEEbiography}[{\includegraphics[width=1in,height=1.25in,clip,keepaspectratio]{Revision_Round3/Authors_Photos/Picture_Wonjae.png}}]{Wonjae Shin} (Senior Member, IEEE) received the B.S. and M.S. degrees from the Korea Advanced Institute of Science and Technology in 2005 and 2007, respectively, and the Ph.D. degree from the Department of Electrical and Computer Engineering, Seoul National University (SNU), South Korea, in 2017. He has been a Visiting Scholar and a Postdoctoral Research Fellow at Princeton University, Princeton, NJ, USA, from 2016 to 2018. From 2007 to 2014, he was a Member of the Technical Staff at the Samsung Advanced Institute of Technology and Samsung Electronics Co., Ltd., South Korea, where he contributed to next-generation wireless communication networks, especially for 3GPP LTE/LTE-advanced standardizations. Since 2023, he has been with the School of Electrical Engineering, Korea University, Seoul, South Korea, where he is currently an Associate Professor. Prior to joining Korea University, he was a Faculty Member at Pusan National University, Busan, and Ajou University, Suwon, South Korea. His research interests include the design and analysis of future wireless communication systems, such as interference-limited networks and machine learning for wireless networks.

Dr. Shin was the recipient of the Fred W. Ellersick Prize and the Asia-Pacific Outstanding Young Researcher Award from the IEEE Communications Society in 2020, the ICTC (International Conference on ICT Convergence) Best Workshop Paper Award in 2022, the Journal of Korean Institute of Communications and Information Sciences (J-KICS) Best Paper Award in 2021, the Best Ph.D. Dissertation Award from SNU in 2017, Gold Prize from the IEEE Student Paper Contest (Seoul Section) in 2014, and the Award of the Ministry of Science and ICT of Korea in IDIS-Electronic News ICT Paper Contest in 2017. He was a co-recipient of the SAIT Patent Award (2010), the Samsung Journal of Innovative Technology Award (2010), the Samsung Human Tech Paper Contest (2010), and the Samsung CEO Award (2013). He was recognized as an Exemplary Reviewer by IEEE WIRELESS COMMUNICATIONS LETTERS in 2014 and IEEE TRANSACTIONS ON COMMUNICATIONS in 2019. He was also awarded several fellowships, including the Samsung Fellowship Program in 2014 and the SNU Long-Term Overseas Study Scholarship in 2016. He is an Associate Editor for the IEEE OPEN JOURNAL OF COMMUNICATIONS SOCIETY.
\end{IEEEbiography}
}
\end{document}